\documentclass{optica-article}

\journal{opticajournal} 

\articletype{Research Article}

\usepackage{lineno}
\usepackage{tabularx}
\usepackage{booktabs}
\usepackage{braket}
\usepackage{dsfont}

\newcommand{\secref}[1]{Sec.~\ref{#1}}
\newcommand{\eqnref}[1]{Eq.~\eqref{#1}}
\newcommand{\figref}[1]{Fig.~\ref{#1}}
\newcommand{\rr}{\mathbf{r}}
\newcommand{\BB}{\mathbf{B}}
\newcommand{\EE}{\mathbf{E}}
\newcommand{\kk}{\mathbf{k}}
\newcommand{\w}{\omega}
\newcommand{\uv}{\mathbf{e}}
\newcommand{\spare}[1]{\left[ {#1} \right]}
\newcommand{\pare}[1]{\left( {#1} \right)}
\newcommand{\cpare}[1]{\left\{ {#1} \right\}}
\newcommand{\abs}[1]{\left |{#1} \right |}

\begin{document}

\title{Quantum Electrodynamics with a Nonmoving Dielectric Sphere: Quantizing Lorenz-Mie Scattering}

\author{Patrick Maurer,\authormark{1,2} Carlos Gonzalez-Ballestero,\authormark{1,2} Oriol Romero-Isart\authormark{1,2,*}}

\address{\authormark{1}Institute for Theoretical Physics, University of Innsbruck, A-6020 Innsbruck, Austria.\\
\authormark{2}Institute for Quantum Optics and Quantum Information of the Austrian Academy of Sciences, A-6020 Innsbruck, Austria.}

\email{\authormark{*}Oriol.Romero-Isart@uibk.ac.at} 


\begin{abstract*} We quantize the electromagnetic field in the presence of a nonmoving dielectric sphere in vacuum. The sphere is assumed to be lossless, dispersionless, isotropic, and homogeneous. The quantization is performed using normalized eigenmodes as well as plane-wave modes. We specify two useful alternative bases of normalized eigenmodes: spherical eigenmodes and scattering eigenmodes. A canonical transformation between plane-wave modes and normalized eigenmodes is derived. This formalism is employed to study the scattering of a single photon, coherent squeezed light, and two-photon states off a dielectric sphere. In the latter case we calculate the second-order correlation function of the scattered field, thereby unveiling the angular distribution of the Hong-Ou-Mandel interference for a dielectric sphere acting as a three-dimensional beam splitter. Our results are analytically derived for a dielectric sphere of arbitrary refractive index and size with a particular emphasis on the small-particle limit. As shown in \cite{Maurer2023}, this work sets the theoretical foundation for  describing the quantum interaction between light and the motional, rotational and vibrational degrees of freedom of a dielectric sphere.
\end{abstract*}

\section{Introduction}
\label{sec:Intro}
The interaction between light and a dielectric sphere is a cornerstone in the theory of electrodynamics and the core topic of current experiments in optical levitated optomechanics~\cite{MillenRepProgPhys2020,GonzalezBallestero2021}. These experiments have demonstrated motional ground-state cooling of an optically levitated nanoparticle~\cite{Delic2020, Magrini2021, Tebbenjohanns2021,Ranfagni2022,Piotrowski2022,Kamba2022}. This long coveted milestone opens the door to controlling the quantum dynamics of a dielectric nanoparticle interacting with light, and to novel research directions in quantum sensing, non-equilibrium quantum physics, and the foundations of quantum mechanics~\cite{GonzalezBallestero2021}. 
It is thus timely to develop a deep theoretical understanding of the interaction between light and a dielectric sphere in the quantum regime. Thus far, the theory used is based on modelling sub-wavelength spheres as point electric dipoles with a polarizability obtained from classical electrostatics \cite{Romero-Isart2010, Chang2010,  Barker2010, Romero-Isart2011, Pflanzer2012, Rodenburg2016,Tebbenjohanns2019,GonzalezBallestero2019, Toros2020,Rudolph2021,Toros2021}. 
Our goal is to develop a theory of quantum electrodynamics in the presence of a dielectric sphere beyond the point-dipole approximation. Its motivation is to provide refined predictions of the quantum dynamics of the nanoparticle (motion, vibrations, rotations) that are relevant to current experiments~\cite{Ashkin1977,Li2011,BlakemorePRA2019,MonteiroPRA2020,KawasakiRSI2020,AfekPRD2021,BlakemorePRD2021,MooreQST2021,Arita2022,PrielSciAdv2022}, and to propose new experiments optically controlling larger particles (e.g. Lorenz-Mie) in the quantum regime.

In this article we focus on deriving quantum electrodynamics in the presence of a rigid and nonmoving dielectric sphere. We quantize the electromagnetic field in the presence of a dielectric sphere in terms of spherical, scattering, and plane-wave modes (\secref{sec:quantization}). We then focus on the problem of quantum light scattering (\secref{sec:qmie}). We study the scattering of single-photon states, coherent squeezed light, and two-photon states. This theoretical framework provides the essential toolbox to further develop the theory by including quantum degrees of freedom of the dielectric sphere (e.g. center-of-mass motion, acoustic vibration) that are at the core of levitated optomechanical experiments. This significant extension is presented in~\cite{Maurer2023} for the particularly relevant case of center-of-mass motion.

This article builds upon the seminal work of R.~J.~Glauber and M.~Lewenstein~\cite{Glauber1991}, where a theory of quantum electrodynamics in the presence of general inhomogeneous and linear dielectric media is derived. Here, we particularize to the case of a dielectric sphere where, by making use of analytical results derivable in the presence of spherical symmetry, some of them very recent~\cite{McPhedran2020},  we significantly extend their results. In order to make our article self-consistent with a coherent notation and with a logical derivation, we re-derive some of the results contained in Ref.~\cite{Glauber1991}. 

\section{Quantization of the electromagnetic field with a dielectric sphere}
\label{sec:quantization}

\begin{figure}[tbh!]
	\centering
	\includegraphics[width=0.7\textwidth]{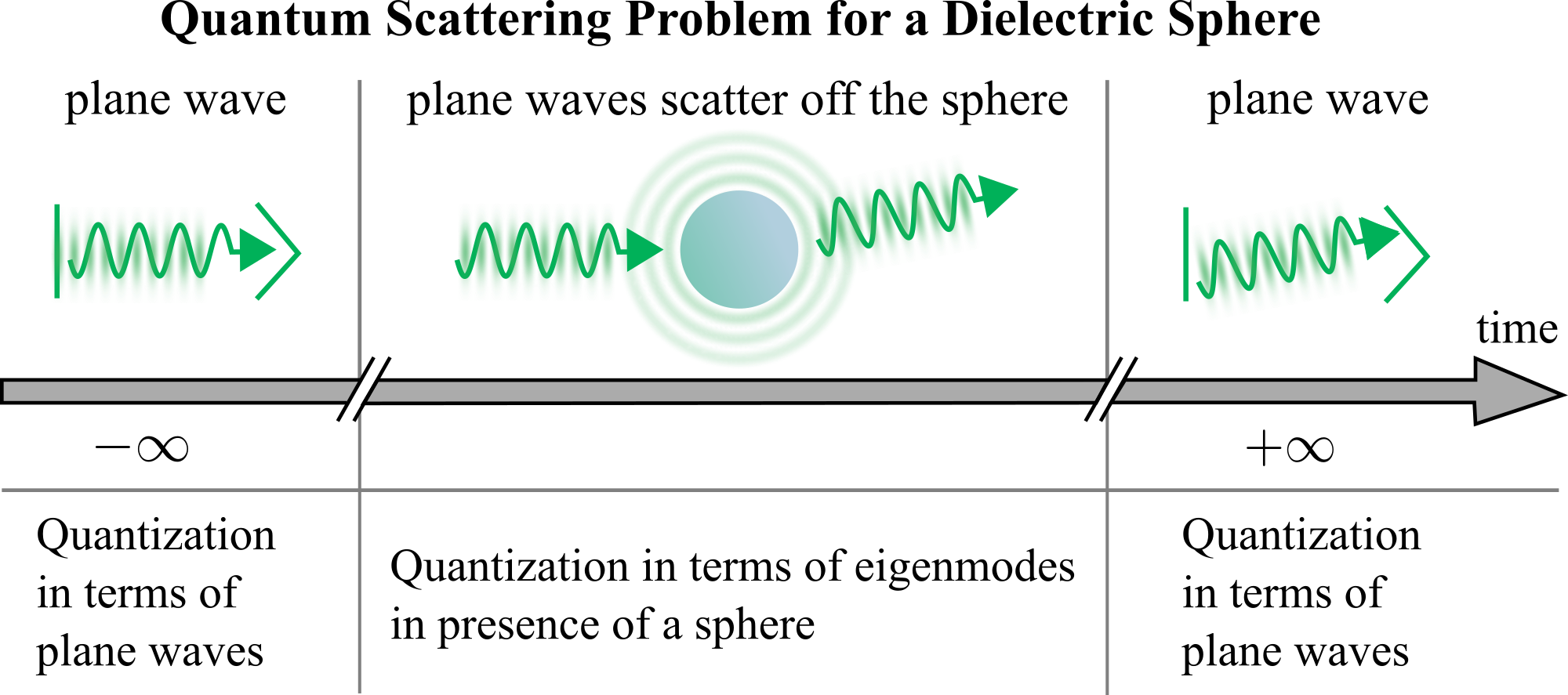}
	\caption{The quantum scattering problem for a dielectric sphere can be solved  by quantizing the electromagnetic field both in terms of plane waves and of eigenmodes. The plane waves describe the asymptotic scattering states, whereas the eigenmodes allow us to derive the single-photon scattering matrix.}\label{FigureBigPicture}
\end{figure}

In this article we use three key ingredients to theoretically study the scattering of quantum electromagnetic states off a dielectric sphere  (see~Fig.~\ref{FigureBigPicture}). (i)~The quantization of the electromagnetic field in terms of plane waves, both in the presence and in the absence of the sphere. Plane waves are the natural initial and final states in scattering problems with localized scatterers. (ii)~The quantization of the electromagnetic field in terms of normalized eigenmodes of Maxwell's equations in the presence of the sphere. As opposed to plane waves (which are not eigenmodes), using normalized eigenmodes will enable us to simplify the scattering problem and formally solve it exactly. Similarly to free space, the normalized eigenmodes are degenerate, i.e. there are multiple eigenmodes with the same eigenvalue (i.e frequency). This allows us to define multiple alternative orthonormal bases of normalized eigenmodes. We exploit this freedom to define two specially convenient bases, namely the symmetric \textit{normalized spherical eigenmodes}, useful in analytical derivations, and the \textit{normalized scattering eigenmodes}, useful to recover the classical Lorenz-Mie scattering results and to compactly express the single-photon scattering matrix. (iii)~A canonical transformation relating creation and annihilation operators of normalized eigenmodes and plane waves. This will ultimately allow us to establish a relation between the quantum electromagnetic states in the distant past and in the distant future (i.e. between the initial and final plane waves), thus solving the scattering problem.

In this section we derive these three key ingredients. First we summarize the classical equations of motion and Hamiltonian of the electromagnetic fields in Sec.~\ref{subsec:ClassicalEOM}. We then define general normalized eigenmodes and summarize the canonical quantization in the presence of a dielectric sphere in Sec.~\ref{sec:normeigenmodes}. We then define the two useful bases of normalized eigenmodes in Secs.~\ref{sec:sphericalmodes} and \ref{sec:scatteringmodes}, respectively. Finally, we derive the quantization in terms of plane-wave modes and the canonical transformation between these and normalized eigenmodes in Sec.~\ref{sec:planwaves}.

\subsection{Preliminaries: classical equations of motion}\label{subsec:ClassicalEOM}

Let us start by describing the classical dynamics of the electric $\mathbf{ E}(\rr,t)$ and magnetic $\mathbf{B}(\rr,t)$ field in the presence of a nonmoving and rigid dielectric sphere of radius $R$ centered at $\rr = \mathbf{0}$. The dielectric sphere is assumed to be lossless, dispersionless, isotropic, and homogeneous with scalar relative permittivity $\epsilon >1$, and surrounded by vacuum ($\epsilon =1$). Taking the permittivity as dispersionless is an approximation used to describe the interaction of a dielectric sphere with monochromatic and narrow-band electromagnetic fields, a common setting in optical levitodynamics~\cite{Romero-Isart2010, Chang2010,  Barker2010, Romero-Isart2011, Tebbenjohanns2019, Delic2020, Magrini2021, Tebbenjohanns2021}. The dielectric medium is thus described by the relative permittivity function
\begin{equation}
\label{eq:epsilon_sphere}
\epsilon(\rr)=1+(\epsilon-1)\Theta(R-|\rr|),
\end{equation}
where $\Theta(x)$ is the Heaviside step function. 
In this medium, the equations of motion of the electromagnetic fields are given by the Maxwell equations
\begin{align}
\nabla\cdot[\epsilon(\rr)\mathbf E(\rr,t)]&=0,\\
\nabla\cdot\mathbf{B}(\rr,t)&=0,\\
\nabla\times\mathbf{E}(\rr,t)+\partial_t\mathbf{B}(\rr,t)&=0,\\
c^2\nabla\times\mathbf{B}(\rr,t)-\epsilon(\rr)\partial_t\mathbf{E}(\rr,t)&=0.
\end{align}
Here $c=1/\sqrt{\epsilon_0 \mu_0}$ is the speed of light in vacuum with $\epsilon_0$ ($\mu_0$) the vacuum permittivity (permeability). 

The electric and magnetic fields can be expressed using the vector potential $\mathbf{A}(\rr,t)$ as 
\begin{align}
\mathbf{E}(\rr,t)&=-\partial_t \mathbf{A}(\rr,t), \\
\mathbf{B}(\rr,t)&=\nabla\times\mathbf{A}(\rr,t).
\end{align}
 The vector potential is defined in the generalized Coulomb gauge
\begin{equation} \label{eq:generalizedgauge}
\nabla\cdot [\epsilon(\rr)\mathbf{A}(\rr,t)]=0.
\end{equation}
Hereafter, we define any vector field $\mathbf{X}(\rr,t)$ as ``$\epsilon$-transverse'' if it fulfills $\nabla\cdot [\epsilon(\rr)\mathbf{X}(\rr,t)]=0$. Similarly, we define it as  ``transverse'' if it fulfills $\nabla\cdot \mathbf{X}(\rr,t)=0$. Hence, note that $\mathbf{A}(\rr,t)$ and $\mathbf{E}(\rr,t)$ are $\epsilon$-transverse while $\mathbf{B}(\rr,t)$ is transverse.

From Maxwell's equations one can derive the following equation of motion for the vector potential $\mathbf{A}(\rr,t)$,
\begin{align}
\label{eq:vector_wave_equation}
\nabla\times\nabla\times \mathbf{A}(\rr,t)+\frac{\epsilon(\rr)}{c^2}\partial_t^2\mathbf{A}(\rr,t)=0.
\end{align}
The electromagnetic equations of motion can be derived from the following classical Hamiltonian~\cite{Cohen2004/1}
\begin{equation}
\label{eq:classcial_hamiltonian}
H=\int_{\mathbb{R}^3}\text{d}\rr \left[\frac{\mathbf{\Pi}^2(\rr,t)}{2\epsilon_0 \epsilon(\rr)}+\frac{[\nabla \times \mathbf{A}(\rr,t)]^2}{2\mu_0}\right],
\end{equation}
where we have defined the transverse conjugate momentum field
\begin{equation}
\mathbf{\Pi}(\rr,t) \equiv \epsilon_0 \epsilon(\rr) \partial_t \mathbf{A}(\rr,t).
\end{equation}
 The Hamilton equations derived from the Hamiltonian \eqnref{eq:classcial_hamiltonian} are given by
\begin{align}
\label{eq:hamilton_vector_potential}
\partial_t \mathbf{A}(\rr,t)&=\frac{\mathbf{\Pi}(\rr,t)}{\epsilon_0 \epsilon(\rr)},\\
\label{eq:hamilton_conjugate_momentum}
\partial_t\mathbf{\Pi}(\rr,t)&=-\frac{1}{\mu_0}\nabla\times\nabla\times\mathbf{A}(\rr,t).
\end{align}
These Hamilton equations are equivalent both to Maxwell equations and to the vector wave equation \eqnref{eq:vector_wave_equation}. 

Canonical quantization of the electromagnetic field requires expressing the fields in terms of normalized eigenmodes of the dynamical equations, with their corresponding eigenfrequencies. In the presence of a dielectric medium, a proper definition of these normalized eigenmodes and their orthogonality relations is key to enable canonical quantization. In the following section we define the normalized eigenmodes in the presence of the dielectric sphere and outline the canonical quantization procedure.

\subsection{Normalized eigenmodes and canonical quantization} \label{sec:normeigenmodes}

Let us introduce a set of complex-valued vector functions $\mathbf{A}_\alpha(\rr)$ with $\alpha$ being 
a multi-index, possibly containing both discrete and continuous indices, that is kept unspecified in this section.  We  define the vector functions $\mathbf{A}_\alpha(\rr)$ as {\em normalized eigenmodes} if they fulfill the following, not necessarily independent, properties:

\begin{enumerate}

\item $\mathbf{A}_\alpha(\rr)$ are solutions of the eigenmode equation
\begin{align}
\label{eq:eigenmode_equation}
\nabla\times\nabla\times \mathbf{A}_\alpha(\rr) - \epsilon(\rr)\frac{ \w_\alpha^2}{c^2}\mathbf{A}_\alpha(\rr)=0.
\end{align}
The variable $\omega_\alpha\in \mathbb{R}$ is the corresponding {\em eigenfrequency} to the eigenmode $\mathbf{A}_\alpha(\rr)$~\cite{Joannopoulos1999}.

\item $\mathbf{A}_\alpha(\rr)$ are $\epsilon$-transverse. This property follows from \eqnref{eq:eigenmode_equation}.

\item $\mathbf{A}_\alpha(\rr)$ are finite in all space and they fulfill the interface conditions for electromagnetic fields in presence of a dielectric sphere, namely
\begin{align}\label{eq:interface}
[\mathbf{A}^>_\alpha(R\mathbf{e}_r)-\epsilon \mathbf{A}^<_\alpha(R\mathbf{e}_r)]\cdot \uv_r&=0,\\
[\mathbf{A}^>_\alpha(R\mathbf{e}_r)- \mathbf{A}^<_\alpha(R\mathbf{e}_r)]\times \uv_r&=0,\\
[\nabla\times\mathbf{A}^>_\alpha(R\mathbf{e}_r)-\nabla\times\mathbf{A}^<_\alpha(R\mathbf{e}_r)]\cdot \uv_r&=0,\\
[\nabla\times\mathbf{A}^>_\alpha(R\mathbf{e}_r)- \nabla\times\mathbf{A}^<_\alpha(R\mathbf{e}_r)]\times \uv_r&=0.
\end{align}
Here $\uv_r\equiv\mathbf{e}_r(\theta,\phi)$ is the standard radial unit vector in spherical coordinates, and $\mathbf{A}^>_\alpha(\rr)$ and $\mathbf{A}^<_\alpha(\rr)$ denote the solutions to the eigenmode equation \eqnref{eq:eigenmode_equation} outside and inside the sphere, respectively~\cite{Jackson1999}. Spherical coordinates are defined in the standard form $\rr = r (\sin \theta \cos \phi, \sin \theta \sin \phi, \cos \theta)$.

\item $\mathbf{A}_\alpha(\rr)$ fulfill the Silver-M\"uller radiation or absorption condition~\cite{Mueller1957,Silver1949,Schot1992},
\begin{align}\label{eq:silver_mueller_condition_0}
    \lim_{r\rightarrow \infty}r \Big\{ \big[ \nabla \times \big(\mathbf{A}_\alpha(\rr)-\lim_{\epsilon\to 1}\mathbf{A}_\alpha(\rr)  \big)\big]\times \uv_r 
    \mp \text{i}(\omega_\alpha/c)  \big(\mathbf{A}_\alpha(\rr)-\lim_{\epsilon\to 1}\mathbf{A}_\alpha(\rr)  \big)  \Big\}=0.
\end{align}
This condition ensures the uniqueness of solutions to the vector Helmholtz equation just as the Sommerfeld radiation and absorption conditions do for solutions to the scalar Helmholtz equation~\cite{Schot1992}. Physically, the integration of Maxwell's equations using Green's theorem allows us to express the electromagnetic fields at any given point within a spatial region $\mathcal{V}\subseteq \mathbb{R}^3$ in terms of the sources within $\mathcal{V}$ and of the electromagnetic fields at its boundaries. The latter terms originate from sources outside of $\mathcal{V}$ and their contributions must vanish for unbounded regions, a requirement manifested mathematically via~\eqnref{eq:silver_mueller_condition_0}. Thus, the Silver-M\"uller condition amounts to assuming no sources at infinity.

\item $\mathbf{A}_\alpha(\rr)$ are orthogonal \cite{Joannopoulos1999} and normalized according to
\begin{equation}
\int_{\mathbb{R}^3}\text{d}\rr \epsilon(\rr)\mathbf{ A}_\alpha^{*}(\rr)\cdot\mathbf{ A}_{\alpha'}(\rr)=\delta_{\alpha\alpha'}.
\end{equation}
Here $\delta_{\alpha\alpha'}$ contains a Kronecker (Dirac) delta for each discrete (continuous) index.

\item The set of vector fields $\mathbf{A}_\alpha(\rr)$ is a complete basis for $\epsilon$-transverse fields, i.e., any $\epsilon$-transverse vector potential $\mathbf{A}(\rr,t)$ and any transverse conjugate momentum field $\mathbf{\Pi}(\rr,t)$ can be expanded as
\begin{align}
\label{eq:expand_vector_potential}
\mathbf{A}(\rr,t)&=\sum_{\alpha}A_\alpha(t)\mathbf{A}_\alpha(\rr) ,\\
\label{eq:expand_conjugate_momentum}
\mathbf{\Pi}(\rr,t)&=\sum_{\alpha}\Pi_\alpha(t)\epsilon(\rr)\mathbf{A}_\alpha(\rr).
\end{align}
Here $\sum_{\alpha}$ includes sums (integrals) over discrete (continuous) indices.
The time-dependent coefficients are given by
\begin{align}
\label{eq:coefficients_vector_potential}
A_\alpha(t)&=\int_{\mathbb{R}^3}\text{d}\rr \epsilon(\rr)\mathbf{A}^*_\alpha(\rr)\cdot\mathbf{A}(\rr,t),\\
\label{eq:coefficients_conjugate_momentum}
\Pi_\alpha(t)&=\int_{\mathbb{R}^3}\text{d}\rr\mathbf{A}^*_\alpha(\rr)\cdot\mathbf{\Pi}(\rr,t).
\end{align}

\end{enumerate}

Any electrodynamic field is determined, given a set of normalized eigenmodes, by the time-dependent coefficients Eqs.~(\ref{eq:coefficients_vector_potential}) and (\ref{eq:coefficients_conjugate_momentum}). Since the expanded fields \eqnref{eq:expand_vector_potential} and~(\ref{eq:expand_conjugate_momentum}) satisfy Hamilton Eqs.~(\ref{eq:hamilton_vector_potential}) and (\ref{eq:hamilton_conjugate_momentum}) it follows that the coefficients $A_\alpha(t)$ and $\Pi_\alpha(t)$ must fulfill the following dynamical equations,
\begin{align}
\epsilon_0 \partial_t  A_\alpha(t)&=\Pi_\alpha(t),\\
\partial_t\Pi_\alpha(t)&=-\epsilon_0 \omega_\alpha^2 A_\alpha(t).
\end{align}
To simplify the above equations, it is convenient to define the {\em normal variables} $a_\alpha(t)$ and $b_\alpha(t)$ through the relation
\begin{align}
A_\alpha(t)& \equiv \sqrt{ \frac{ \hbar }{2 \epsilon_0\omega_\alpha}}  [a_\alpha(t)-b_\alpha(t)],\\
\Pi_\alpha(t)& \equiv -\text{i} \sqrt{\frac{\epsilon_0 \hbar \omega_\alpha }{2}} [a_\alpha(t)+b_\alpha(t)].
\end{align}
The prefactor including the reduced Planck constant $\hbar$ is chosen here for later convenience in performing canonical quantization. In terms of the normal variables, Hamilton's equations simplify to the single equation
\begin{equation}
 \partial_t a_\alpha(t)=-\text{i} \omega_\alpha a_\alpha(t).
\end{equation}
This is a consequence of the normal variables $b_\alpha(t)$ not being independent, as they can be obtained from $a_\alpha(t)$ via the relation
\begin{equation} \label{eq:normalvariableb}
b_{\alpha}(t)=-\sum_{\alpha'} M_{\alpha \alpha'}a^*_{\alpha'}(t),
\end{equation}
where
\begin{align} \label{eq:Mmatrix}
M_{\alpha\alpha'} = M_{\alpha'\alpha}= \int_{\mathbb{R}^3}\text{d}\rr \epsilon(\rr) \mathbf{A}_\alpha^*(\rr)\cdot\mathbf{A}^*_{\alpha'}(\rr).
\end{align}
This can readily be shown by realizing that, since $\omega_\alpha$ and $\epsilon(\rr)$ are real, $\mathbf{A}^*_\alpha(\rr)$ is also an eigenmode with the same eigenfrequency $\omega_\alpha$ that can be expressed as $\mathbf{A}^*_{\alpha}(\rr)=\sum_{\alpha'} M_{\alpha\alpha'} \mathbf{A}_{\alpha'}(\rr)$, with $M_{\alpha\alpha'}\propto\delta(\omega_\alpha-\omega_{\alpha'})$. Consequently, since $\mathbf{A}(\rr,t)$ and $\mathbf{\Pi}(\rr,t)$ are real it follows that $A_\alpha(t)=\sum_{\alpha'} M_{\alpha\alpha'}A^*_{\alpha'}(t)$ and $\Pi_\alpha(t)=\sum_{\alpha'} M_{\alpha \alpha'}\Pi^*_{\alpha'}(t)$. These expressions along with the definition of the normal variables lead to \eqnref{eq:normalvariableb}. 

The vector potential \eqnref{eq:expand_vector_potential} and  the conjugate momentum \eqnref{eq:expand_conjugate_momentum} are written in terms of the normal variables $a_\alpha(t)$ as
\begin{align}
\mathbf{A}(\rr,t)&=\sum_\alpha \sqrt{ \frac{ \hbar }{2 \epsilon_0\omega_\alpha}}[a_\alpha(t)\mathbf{A}_\alpha(\rr)+\text{c.c.}],\\
\mathbf{\Pi}(\rr,t)&=- \text{i} \sum_\alpha \sqrt{\frac{\epsilon_0 \hbar \omega_\alpha }{2}} [a_\alpha(t)\epsilon(\rr)\mathbf{A}_\alpha(\rr)-\text{c.c.}].
\end{align}
By inserting these expressions into \eqnref{eq:classcial_hamiltonian}, the classical Hamiltonian can also be expressed in terms of the normal variables as
\begin{equation} \label{eq:classicalHnormalvariables}
H=\hbar \sum_\alpha \frac{\omega_\alpha}{2} [a_\alpha^*(t)a_\alpha(t)+a_\alpha(t)a^*_\alpha(t)].
\end{equation}

With the normal variables and normalized eigenmodes properly defined, canonical quantization can now be readily performed. Normal variables are replaced by creation and annihilation operators, namely $a_\alpha \rightarrow \hat a_\alpha$ and  $a^*_\alpha \rightarrow \hat a^\dagger_\alpha$, which fulfill the bosonic commutation rules $[\hat{a}_\alpha,\hat{a}_{\alpha'}^\dagger]=\delta_{\alpha\alpha'}$ and $[\hat{a}_\alpha,\hat{a}_{\alpha'}]=[\hat{a}^\dagger_\alpha,\hat{a}_{\alpha'}^\dagger]=0$. The Hamilton operator is then given by
\begin{equation}
\label{eq:hamilton_operator_eigenmodes}
\hat{H}=\hbar \sum_\alpha \omega_\alpha \pare{\hat{a}_\alpha^\dagger \hat{a}_\alpha +\frac{1}{2}},
\end{equation}
and the electric and magnetic field operators by
\begin{align}
\label{eq:electric_operator_eigenmodes}
\hat{\mathbf{E}}(\rr)&=\text{i}\sum_{\alpha}\sqrt{\frac{\hbar \omega_\alpha}{2\epsilon_0}}\spare{\mathbf{A}_\alpha(\rr)\hat{a}_\alpha-\text{H.c.}},\\
\label{eq:magnetic_operator_eigenmodes}
\hat{\mathbf{B}}(\rr)&=\sum_{\alpha}\sqrt{\frac{\hbar}{2\epsilon_0\omega_\alpha}}\spare{\nabla\times\mathbf{A}_\alpha(\rr)\hat{a}_\alpha+\text{H.c.}}.
\end{align}

In this section we have deliberately kept the particular form of the normalized eigenmodes unspecified. The above theoretical treatment of the classical and quantum electromagnetic field in the presence of a dielectric sphere thus applies to any set of normalized eigenmodes $\mathbf{A}_\alpha(\rr)$. This set is not unique, i.e. one can express the fields in different bases of normalized eigenmodes. A simple example is the electromagnetic field in free space, which can be expressed in terms of plane wave eigenmodes and spherical wave eigenmodes. This freedom ultimately stems from the degeneracy of the eigenmodes, since within the subspace of eigenmodes with the same eigenfrequency, an arbitrary orthonormal basis can be chosen. In the presence of a dielectric sphere, we will use this freedom to our advantage and define two alternative bases of normalized eigenmodes, each convenient at different stages in the scattering problem. In the next two sections we introduce these two bases of normalized eigenmodes.

\begin{table}[t]
	\centering
	\begin{tabular}{ p{2.2cm}  p{2.2cm}  p{2.2cm}  p{1.6cm} }
		
		&$\mathbf{S}_\alpha(\rr)$ & $\mathbf{F}_\kappa(\rr)$ & $\mathbf{G}_\kappa(\rr)$  \\
		\toprule
		eigenmodes &  yes (spherical)   & yes (scattering) &  no \\
		\midrule
		normalized & yes & yes & yes \\
		\midrule
		transversality & $\epsilon$-transverse& $\epsilon$-transverse & transverse\\
		\midrule
		indices &  $\alpha = (p,l,m,k)$  &  $\kappa = (g,\kk)$  &  $\kappa = (g,\kk)$ \\ 
		&  $p\in \lbrace \text{TE, TM}\rbrace$  &  $g\in \lbrace1,2\rbrace$  &  $g\in \lbrace1,2\rbrace$ \\
		&  $|m|<l \in \mathbb{N} $  &  $\kk \in \mathbb{R}^3$  &   $\kk \in \mathbb{R}^3$  \\
		&  $m\in \mathbb{Z}$, $k \in \mathbb{R}$  &  &  \\
		\midrule
		operators & $\hat{a}_\alpha$ & $\hat{a}_\kappa$ &  $\hat{b}_\kappa$\\
		\midrule
		definitions & Eqs.\eqref{eq:spherical_eigenmodes1}-\eqref{eq:spherical_eigenmodes2} &  \eqnref{eq:scattering_eigenmodes} & \eqnref{eq:plane_wave}\\
		\bottomrule	
	\end{tabular}
	\caption{Summary of the most important properties of all the families of modes and eigenmodes employed across the text.} \label{tab:1}
\end{table}

\subsection{Normalized Spherical Eigenmodes}
\label{sec:sphericalmodes}

The first basis of normalized eigenmodes, which we name {\em normalized spherical eigenmodes}, are the modes resulting from solving the eigenmode equation in spherical coordinates. Physically these normalized eigenmodes are related to the multipolar expansion of the fields in the presence of the sphere, as shown below. Aside from being useful for analytical derivations, these modes are a necessary ingredient in the definition of the second family of eigenmodes (see \ref{sec:scatteringmodes}), which in turn are used to study scattering. We denote the normalized spherical eigenmodes by $\mathbf{S}_\alpha (\rr)$.

\subsubsection{Derivation}

To derive the normalized spherical eigenmodes, we first obtain the solutions to the eigenmode equation inside and outside the sphere, and then match them using the boundary conditions. We start by considering the vector functions $\mathbf{N}^\epsilon_\alpha (\rr)$ that solve the vector Helmholtz equation for a \textit{homogeneous} dielectric medium with relative permittivity $\epsilon(\rr)=\epsilon$, namely 
\begin{equation}
\label{eq:spherical_vector_helmholtz}
\nabla\times\nabla\times \mathbf{N}^\epsilon_\alpha(\rr) - \epsilon \frac{ \omega_\alpha^2}{c^2}\mathbf{N}^\epsilon_\alpha(\rr)=0.
\end{equation}
These functions can be constructed from the
solutions of the scalar Helmholtz equation in spherical coordinates,
\begin{equation}
\nabla^2 \psi^\epsilon_{lm}(k;\rr)  +\epsilon\frac{\omega_\alpha^2}{c^2}\psi^\epsilon_{lm}(k;\rr)=0,
\end{equation}
which are given by
\begin{equation}
\label{eq:spherical_scalar_helmholtz}
\psi^\epsilon_{lm}(k;\rr)  = A^\epsilon_{lm}(k) j_l(\sqrt{\epsilon}k r) Y_l^m(\theta,\phi) \\
+B^\epsilon_{lm}(k) y_l(\sqrt{\epsilon}k r)Y_l^m(\theta,\phi).
\end{equation}
 We have introduced the mode indices $k= \omega_\alpha/c \in \mathbb{R}$, $l\in\mathbb{N}_0$, $m\in\mathbb{Z}$ and $|m|\leq l$. The function $j_l(x)$ ($y_l(x)$) is the spherical Bessel function of the first (second) kind and order $l$, $Y_l^m(\theta,\phi)$ are spherical harmonics, and $A^\epsilon_{lm}(k), B^\epsilon_{lm}(k) \in \mathbb{C}$ are constants. From \eqnref{eq:spherical_scalar_helmholtz} one can construct all the solutions of \eqnref{eq:spherical_vector_helmholtz} using two sets of orthogonal solutions, labelled by the polarization index $p \in \cpare{\text{TE},
\text{TM}}$ that are given by
\begin{equation}
\mathbf{N}^{ \epsilon\text{TE}}_{lm}(k;\rr)\equiv \frac{\text{i} \nabla \times [\rr \psi^\epsilon_{lm} (k;\rr)]}{\sqrt{l(l+1)}}  \\ = g_{llm}^{\epsilon \text{TE}}(k;r)\mathbf{X}_l^m(\theta,\phi),
\end{equation}
and
\begin{align}
\mathbf{N}^{\epsilon \text{TM}}_{lm}(k;\rr)&\equiv -\frac{\sqrt{2l+1}\nabla \times \nabla \times [\rr \psi^\epsilon_{lm}(k;\rr)] }{\sqrt{l(l+1)}\sqrt{\epsilon}k}\\
&=  \sqrt{l}g^{\epsilon\text{TM}}_{ll+1m}(k;r)\mathbf{V}_l^m(\theta,\phi)- \sqrt{l+1}g_{ll-1m}^{\epsilon \text{TM}}(k;r)\mathbf{W}_l^m(\theta,\phi).
\end{align}
Here we have defined the radial function
\begin{equation}
g^{\epsilon p}_{ll'm}(k;r) \equiv A^{\epsilon p}_{lm}(k) j_{l'}(\sqrt{\epsilon}k r)+ B^{\epsilon p}_{lm}(k) y_{l'}(\sqrt{\epsilon}k r),
\end{equation}
with constants $A^{\epsilon p}_{lm} (k), B^{\epsilon p}_{lm}(k) \in  \mathbb{C}$, and the three vector spherical harmonics \cite{Hill1954}
\begin{align}
\label{eq:vector_spherical_harmonics}
\mathbf{X}_l^m(\theta,\phi)& \equiv \frac{\rr\times \nabla Y_l^m(\theta,\phi)}{\text{i}\sqrt{l(l+1)}},\\
\mathbf{V}_l^m(\theta,\phi)& \equiv \frac{-(l+1)Y_l^m(\theta,\phi)\uv_r+r\nabla Y_l^m(\theta,\phi)}{\sqrt{(l+1)(2l+1)}},\\
\mathbf{W}_l^m(\theta,\phi)& \equiv \frac{lY_l^m(\theta,\phi)\uv_r+r\nabla Y_l^m(\theta,\phi)}{\sqrt{l(2l+1)}}.
\end{align}
Each of the solutions is fully determined by the multi-index $\alpha = (p,l,m,k)$, whose physical interpretation is given below. Note also that the solutions corresponding to $l=0$ vanish and therefore hereafter $l\in\mathbb{N}$. 

The normalized spherical eigenmodes $\mathbf{S}_\alpha (\rr)$ are constructed by combining the vector functions $\mathbf{N}^\epsilon_\alpha(\rr)$ inside the sphere ($r \le R$, $\epsilon > 1$) and $\mathbf{N}^1_\alpha (\rr)$ outside the sphere ($r>R$, $\epsilon=1$), and by fixing the four constants $A^{\epsilon p}_{lm}(k), B^{\epsilon p}_{lm}(k), A^{1 p}_{lm}(k), B^{1 p}_{lm}(k) \in  \mathbb{C}$. 

First, we set $B^{\epsilon p}_{lm}(k) =0$ in order to guarantee that $\mathbf{S}_\alpha (\rr)$ is finite when $r \rightarrow 0$. Then, two of the remaining three constants are determined by the interface boundary conditions that all normalized eigenmodes fulfill by definition, namely Eqs.~(\ref{eq:interface}).
 After imposing such boundary conditions we obtain the following (not yet normalized) spherical eigenmodes:
\begin{align}
\mathbf{S}^\text{TE}_{lm}(k;\rr)&=  g_{llm}^\text{TE}(k;r)\mathbf{X}_l^m(\theta,\phi),\\
\mathbf{S}^\text{TM}_{lm}(k;\rr)&= \sqrt{l}g_{ll+1m}^\text{TM}(k;r)\mathbf{V}_l^m(\theta,\phi)- \sqrt{l+1}g_{ll-1m}^\text{TM}(k;r)\mathbf{W}_l^m(\theta,\phi),
\end{align}
where the radial function is now given by
\begin{equation}
g^p_{ll'm}(k;r) \equiv A^{p}_{lm}(k)
\begin{cases}
 j_{l'}(\sqrt{\epsilon}kr ) &  r\leq R,\\
\alpha_l^p j_{l'}(kr)+\beta_l^p y_{l'}(kr) & r> R.
\end{cases}
\end{equation}
Here $A^{p}_{lm}(k) \in  \mathbb{C}$ is an undetermined constant whereas the coefficients $\alpha^{p}_{l} , \beta^{p}_{l}  \in  \mathbb{R}$ are given by 
\begin{align}
\label{eq:boundary_maxwell_coefficients}
\alpha_l^\text{TE}& \equiv q q' j_{l+1}(q')y_l(q)-q^2 j_l(q')y_{l+1}(q),\\
\beta_l^\text{TE}&\equiv q^2 j_{l}(q')j_{l+1}(q)-q q'j_{l+1}(q')j_{l}(q),\\
\alpha_l^\text{TM}&\equiv q^2 j_{l+1}(q')y_{l}(q)-q q' j_l(q')y_{l+1}(q)+q' [\left(\epsilon-1\right)/\epsilon](l+1)j_l(q')y_l(q),\\
\beta_l^\text{TM}&\equiv q q' j_l(q')j_{l+1}(q)-q^2 j_{l+1}(q')j_l(q)-q' [\left(\epsilon-1\right)/\epsilon](l+1)j_l(q')j_l(q),
\end{align}
with the adimensional parameters
\begin{equation}\label{eq:definitionQandQprime}
    q \equiv kR,\hspace{0.8cm} q' \equiv  \sqrt{\epsilon}k R.
\end{equation}

The last undetermined constant $A^{p}_{lm}(k) \in  \mathbb{C}$ is fixed, up to a phase, by the normalization condition
\begin{equation}
\int_{\mathbb{R}^3}\text{d}\rr \epsilon(\rr) \mathbf{ S}^{p*}_{lm}(k;\rr)\cdot\mathbf{ S}^{p'}_{l'm'}(k';\rr)
=\delta_{pp'} \delta_{ll'}\delta_{mm'}\delta(k-k').
\end{equation}
Since the vector spherical harmonics are mutually orthonormal \cite{Hill1954}, the normalization condition simplifies to
\begin{align}
\int_0^\infty \text{d}r r^2 \epsilon(\rr)& g_{llm}^{\text{TE}*}(k;\rr)g_{llm}^\text{TE}(k';\rr)=\delta(k-k'),\\
\notag
\int_0^\infty \text{d}r r^2 \epsilon(\rr)& [ l g_{ll+1m}^{\text{TM}*}(k;\rr)g_{ll+1m}^\text{TM}(k';\rr)\\
+&(l+1) g_{ll-1m}^{\text{TM}*}(k;\rr)g_{ll-1m}^\text{TM}(k';\rr)]=\delta(k-k').
\end{align}
These integrals can be analytically evaluated using the techniques recently reported in~\cite{McPhedran2020}. They lead to
\begin{align}
A^\text{TE}_{lm}(k)&=\sqrt{\frac{2k^2}{\pi}}\gamma^\text{TE}_l\exp[\text{i} \Delta_{lm}^{\text{TE}}(k)],\\
A^\text{TM}_{lm}(k)&=\sqrt{\frac{2k^2}{\pi(2l+1)}}\gamma^\text{TM}_l\exp[\text{i} \Delta_{lm}^\text{TM}(k)],
\end{align}
where 
\begin{equation}
\gamma^p_l \equiv \frac{1}{\sqrt{(\alpha_l^p)^2+(\beta_l^p)^2}}
\end{equation}
 and $\Delta_{lm}^p(k)\in \mathbb{R}$ is an arbitrary phase yet to be fixed. Up to such phase, the normalized spherical eigenmodes $\mathbf{S}_\alpha(\rr) = \mathbf{S}^p_{lm}(k;\rr)$ take the form
\begin{align}
\label{eq:spherical_eigenmodes1}
\mathbf{S}^\text{TE}_{lm}(k;\rr)&= \sqrt{\frac{2k^2}{\pi}} f_{llm}^\text{TE}(k;r)\mathbf{X}_l^m(\theta,\phi),\\
\notag
\mathbf{S}^\text{TM}_{lm}(k;\rr)&=\sqrt{\frac{2k^2}{\pi}} \sqrt{\frac{l}{2l+1}}f_{ll+1m}^\text{TM}(k;r)\mathbf{V}_l^m(\theta,\phi)\\
\label{eq:spherical_eigenmodes2}
&-\sqrt{\frac{2k^2}{\pi}} \sqrt{\frac{l+1}{2l+1}}f_{ll-1m}^\text{TM}(k;r)\mathbf{W}_l^m(\theta,\phi),
\end{align}
where

\begin{equation} 
\label{eq:radial_function_spherical_eigenmodes}
f^p_{ll'm}(k;r) \equiv \exp[\text{i} \Delta^p_{lm}(k)]
\times\begin{cases}
\gamma^p_lj_{l'}(\sqrt{\epsilon}kr ) &  r\leq R,\\
\cos(\varphi_l^p)j_{l'}(kr)+\sin (\varphi_l^p)y_{l'}(kr) & r> R,
\end{cases}
\end{equation}
and we have used the definition
\begin{align}
\cos \varphi_l^p \equiv &\gamma^p_l\alpha^p_l =\frac{\alpha^p_l}{\sqrt{(\alpha_l^p)^2+(\beta_l^p)^2}}, \\
 \sin \varphi_l^p \equiv &\gamma^p_l\beta^p_l = \frac{\beta^p_l}{\sqrt{(\alpha_l^p)^2+(\beta_l^p)^2}}.
\end{align}

As a final step, we specify the phase $\Delta_{lm}^p(k)\in \mathbb{R}$ using the Silver-M\"uller boundary condition~\eqref{eq:silver_mueller_condition_0}. First, we define the normalized spherical eigenmodes in free space, namely in the absence of the sphere, as
\begin{equation}
 \mathbf{S}^0_\alpha(\rr) \equiv \lim_{\epsilon \rightarrow 1 }\mathbf{S}_\alpha(\rr).
 \end{equation}
  The free-space solution $ \mathbf{S}^0_\alpha(\rr)$ has the same form as \eqnref{eq:spherical_eigenmodes1} with a radial function given by $f^{0p}_{ll'm} (k;r) = \exp[\text{i} \lim_{\epsilon \rightarrow 1} \Delta^p_{lm}(k)] j_{l'}(kr)$, and we set $\lim_{\epsilon\to 1} \Delta_{lm}^p(k) = 0$ without loss of generality. 
  Hereafter, we define the \textit{scattered part} of the spherical normalized eigenmode as 
  \begin{equation} \label{eq:defineScatteredPart}
  \mathbf{S}^\text{sc}_\alpha(\rr) \equiv \mathbf{S}_\alpha(\rr) - \mathbf{S}^0_\alpha(\rr),
  \end{equation}
  i.e. as the normalized spherical eigenmode minus its free-space limit.
  The Silver-M\"uller boundary condition can then be expressed in terms of the scattered part as 
  \begin{equation}
\label{eq:silver_mueller_condition}
\lim_{r\rightarrow \infty}r \cpare{ \spare{ \nabla \times \mathbf{S}^\text{sc}_\alpha(\rr)  }\times \uv_r \mp \text{i} k \mathbf{S}^\text{sc}_\alpha(\rr)  }=0.
\end{equation}
Using the asymptotic expressions of the Bessel functions, namely $\lim_{r \rightarrow \infty} j_{l}(kr)= (kr)^{-1}\sin(kr-l\pi/2) $ and $\lim_{r \rightarrow \infty} y_{l}(kr)= - (kr)^{-1} \cos(kr-l\pi/2)$, one can reduce \eqnref{eq:silver_mueller_condition} to
\begin{equation}
1-\exp[\text{i} (\Delta_{lm}^p(k)\pm\varphi_l^p)]=0,
\end{equation}
which can be solved by $\Delta_{lm}^p(k) = \mp \varphi^p_l$ . Each of the two solutions defines a basis of normalized spherical eigenmodes. 
We name the normalized spherical eigenmodes as 
\begin{equation}
\Delta_{lm}^p(k) = 
\begin{cases}
 - \varphi^p_l &  \text{``outgoing''} \longrightarrow \mathbf{S}^\text{out}_\alpha(\rr),\\
 \varphi^p_l & \text{``incoming''} \longrightarrow \mathbf{S}^\text{in}_\alpha(\rr).
\end{cases}
\end{equation}
This distinction will become relevant for the scattering problem. Note that for each particular problem the choice between outgoing and incoming modes is arbitrary, as both form a complete basis in which any $\epsilon-$transverse field can be expanded. 

The final expression for the normalized spherical eigenmodes is analogous to \eqref{eq:spherical_eigenmodes1}, now with an $m-$independent radial function given by
\begin{equation}
f_{ll'}^{\text{out}p}(k;r) =\\
    \begin{cases}
     \exp(-\text{i} \varphi_l^p)\gamma_l^pj_{l'}(\sqrt{\epsilon}kr) &  r\leq R,\\
     j_{l'}(kr)-\text{i} \sin\varphi_l^p\exp(- \text{i} \varphi_l^p)h_{l'}(kr) & r > R,
    \end{cases}
\end{equation}
and $f_{ll'}^{\text{in}p}(k;r)=[f_{ll'}^{\text{out}p}(k;r)]^*$, where the spherical Hankel function of the first kind and order $l$ reads $h_l(x)\equiv j_l(x) + \text{i} y_l(x).$
The asymptotic limit of these functions is given by

\begin{equation}\label{eq:ffradial}
\lim_{r \rightarrow \infty}f^{\text{out}p}_{ll'}(k;r) =  j_{l'}(kr) 
-(-\text{i})^{l'}\sin(\varphi_l^p)\frac{\exp[ \text{i} (kr - \varphi_l^p)]}{kr},
\end{equation} 
and 
\begin{equation}
\lim_{r \rightarrow \infty}f^{\text{in}p}_{ll'}(k;r) = j_{l'}(kr) 
-\text{i}^{l'}\sin(\varphi_l^p)\frac{\exp[- \text{i} (kr - \varphi_l^p)]}{kr}.
\end{equation}
In the asymptotic limit $\mathbf{S}^\text{out}_\alpha(\rr)$ ($\mathbf{S}^\text{in}_\alpha(\rr)$) therefore corresponds to a linear combination of the vacuum solution $\mathbf{S}^0_\alpha(\rr)$ and an outgoing (incoming) spherical wave. From the above expressions we can identify the mode index $k$ as the modulus of the wavevector of the spherical wave. 
We remark that in  absence of the sphere, since $\lim_{\epsilon \rightarrow 1} \varphi^p_l = 0$, one has that $\mathbf{S}^0_\alpha(\rr) = \lim_{\epsilon \rightarrow 1} \mathbf{S}^\text{out}_\alpha(\rr) = \lim_{\epsilon \rightarrow 1} \mathbf{S}^\text{in}_\alpha(\rr)$.
This concludes our derivation and full characterization of the normalized spherical eigenmodes $\mathbf{S}_\alpha(\rr)$ with $\alpha=(p,l,m,k).$

\subsubsection{Physical insights: multipole expansion and small-particle limit} \label{sec:multipolar}

The normalized spherical eigenmodes $\mathbf{S}_\alpha(\rr)$ are closely related to the electric and magnetic multipole fields of order $l\text{ and }m$. Multipole fields are solutions to the vector Helmholtz equation in vacuum, i.e. in a source-free region of empty space. They are used to expand the electromagnetic field in a source-free region with expansion coefficients that depend on the charge and current distributions~\cite{Jackson1999}. 

Let us now show how the spherical eigenmodes are related to multipole fields. Outside the dielectric sphere, the spherical eigenmodes are in fact defined in terms of solutions to the vector Helmholtz equation in empty space. The electromagnetic fields associated to an eigenmode are given by
\begin{align}\label{eq:modalfield1}
    \EE_\alpha (\rr,t)&=\mathfrak{Re}[\text{i} \omega \mathbf{S}_\alpha(\rr)\exp(-\text{i} \omega t)],\\
    \label{eq:modalfield2}
    \BB_\alpha (\rr,t)&=\mathfrak{Re}[\nabla\times \mathbf{S}_\alpha(\rr)\exp(-\text{i} \omega t)].
\end{align}
The defining property of an electric multipole field of order $l$ and $m$ is that its magnetic field satisfies $\BB_{lm}(\rr) \propto \mathbf{X}_l^m(\theta,\phi)$, and thus $\BB_{lm}(\rr)\cdot\mathbf{e}_r=0$~\cite{Jackson1999}. Conversely, the defining property of a magnetic multipole field of order $l$ and $m$ is that its electric field satisfies $\EE_{lm}(\rr) \propto \mathbf{X}_l^m(\theta,\phi)$, and thus $\EE_{lm}(\rr)\cdot\mathbf{e}_r=0$. Let us now combine Eqs.~\eqref{eq:modalfield1}-\eqref{eq:modalfield2} with the expressions of the normalized spherical eigenmodes Eqs.~\eqref{eq:spherical_eigenmodes1}-\eqref{eq:spherical_eigenmodes2}, and use that both $\nabla\times[f(r)\mathbf{V}_{l}^m(\theta,\phi)]$ and  $\nabla\times[f(r)\mathbf{W}_{l}^m(\theta,\phi)]$ are directly proportional to $\mathbf{X}_l^m(\theta,\phi)$ for arbitrary functions $f(r)$~\cite{Hill1954}. This allows us to conclude that, outside the sphere, the normalized spherical eigenmodes with $p=\text{TM}$ (transverse magnetic field) are proportional to the vacuum electric multipole fields of order $l$ and $m$ whereas those with $p=\text{TE}$ (transverse electric field) are proportional to the vacuum magnetic multipole fields of order $l$ and $m$.

\begin{figure}
	\centering
	\includegraphics[width=0.5\textwidth]{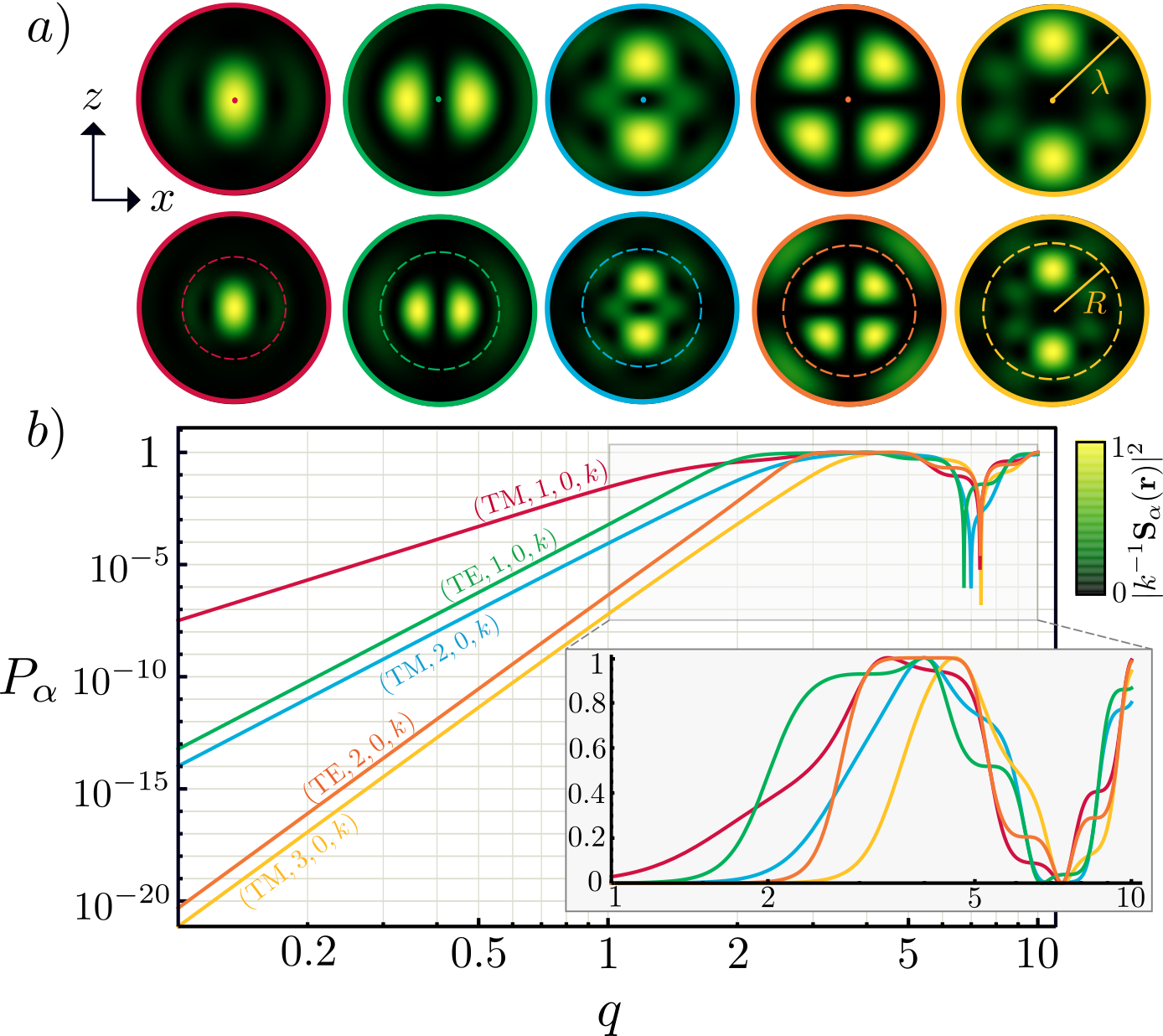}
	\caption{(a) Adimensional squared amplitude $|k^{-1} \mathbf{S}_\alpha(\rr)|^2$ of normalized spherical eigenmodes with $\epsilon=2.1$ in the x-z-plane. From left to right, both rows correspond to the modes $(\text{TM},1,0,k)$, $(\text{TE},1,0,k)$, $(\text{TM},2,0,k)$, $(\text{TM},2,0,k)$, $(\text{TM},3,0,k)$. The upper row shows the squared amplitude for $q=0.1$ (small-particle limit), inside circles of sizes given by the corresponding wavelength $\lambda\equiv 2\pi/k$. In the upper row, the sphere is indicated by the dot at the center of each circle. The lower row shows the squared amplitude at a value of $q$ for which $P_\alpha$ (shown in panel (b)) is maximal, i.e. $q=3.4, 4.0, 4.0, 4.3, 4.6$, respectively. The space occupied by the sphere is indicated by the dashed circle. (b) $P_\alpha$ (\eqnref{eq:integratedIntensity}), as a function of $q$ for the same normalized spherical eigenmodes shown in panel (a). The inset shows the same function in the region $1 \leq q \leq 10$.}
	\label{fig:multipole}
\end{figure}

It is insightful to quantify the adimensional squared amplitude $|k^{-1} \mathbf{S}_\alpha(\rr)|^2$ of the normalized spherical eigenmodes in the vicinity of the sphere. This quantity is relevant for future estimates of interaction between the electromagnetic field and the mechanical degrees of freedom of the sphere, such as motion or acoustic deformation. In \figref{fig:multipole}(a) we display the squared amplitude for some of the five lowest-order modes at the plane $\phi=0$ (see figure's caption for details). The first row of images shows the small-particle limit $q,q' \ll 1$. In this limit, the squared amplitude is negligible near and inside the sphere except for the electric dipole eigenmode $(p=\text{TM},l=1,|m|\leq 1)$. This can be understood from the analytical expression of the eigenmodes by taking the small-particle limit $q,q' \ll 1$  and using the asymptotic expansions $\gamma_l^p j_l(\sqrt{\epsilon} q)= \mathcal{O}(q^l)$ and
\begin{align}\label{eq:smallPlimitPhi}
    \sin \varphi_l^p&=\begin{cases}
     \mathcal{O}(q^{2l+3})&\text{    for    }p=\text{TE},\\
      \mathcal{O}(q^{2l+1})&\text{    for    }p=\text{TM}.
    \end{cases}
\end{align}
The second row of \figref{fig:multipole}(a) shows the squared amplitude of the same modes beyond the small-particle limit. Here, this adimensional quantity inside and near the sphere is comparable for many modes. The results of \figref{fig:multipole}(a) anticipate that, as expected, the interaction between light and the mechanical degrees of freedom of the sphere, which is determined by these profiles, will be dominated by the electric dipole mode in the small-particle limit but will have, for larger particles, a more complex form involving multiple spherical eigenmodes.

As a final discussion, let us relate our solutions to the point-dipole approximation used for scattering problems in the small-particle limit. We start by noting that, in the presence of the sphere, the spherical eigenmodes can be cast as $\mathbf{S}_\alpha(\rr)=\mathbf{S}^0_\alpha(\rr)+\mathbf{S}^\text{sc}_\alpha(\rr)$, see \eqnref{eq:defineScatteredPart}. Thus, the sphere can be considered as the source of the scattered part $\mathbf{S}^\text{sc}_\alpha(\rr)$. This scattered part quantifies how much a vacuum spherical eigenmode is perturbed by the presence of the sphere. 
Usually, in scattering problems only the far-field limit of this perturbation is relevant. In addition, in the far-field (and in general anywhere outside the sphere, as discussed above) the angular dependence of the modes $\mathbf{S}_\alpha(\rr)$ and $\mathbf{S}^0_\alpha(\rr)$ is the same. Hence, we can quantify the sphere-induced perturbation by the integral of $|\mathbf{S}^{\text{sc}}_\alpha(\rr)|^2$ over the solid angle in the far-field, that is
\begin{align}\label{eq:integratedIntensity}
    P_\alpha \equiv \frac{\pi}{2}\lim_{r\rightarrow\infty}r^2\int_{\mathbb{S}^2}\text{d}\Omega |\mathbf{S}^{\text{sc}}_\alpha(\rr)|^2&=\sin^2 \varphi_l^p,
\end{align}
where the analytical expression has been obtained using the orthonormality relations of the vector spherical harmonics together with \eqnref{eq:ffradial}. In \figref{fig:multipole}(b) we display $P_\alpha$ as a function of $q$ and for the first five electric and magnetic multipole fields shown in panel (a). We see that this quantity strongly depends on the order $l$ in the small-particle limit $q,q' \ll 1$ (see Eq.~\eqref{eq:definitionQandQprime}), and is the largest for the electric dipole mode $(p=\text{TM},l=1,|m|\leq 1)$. The decay of $P_\alpha$ with increasing $l$ is exponential, as explicitly shown by Eq.~\eqref{eq:smallPlimitPhi}. 
While in the small-particle limit $(q\ll 1)$ $P_\alpha$ is a monotonously increasing function in $q$, it exhibits minima and maxima in the region $(q>1)$. These extrema amount to the well-known Mie resonances that persist until the optical regime $(q\gg 1)$~\cite{Bohren2004}.

\subsection{Normalized Scattering Eigenmodes}
\label{sec:scatteringmodes}

The sets of incoming or outgoing normalized spherical eigenmodes are not the only bases of normalized eigenmodes that can be used to decompose $\epsilon-$transverse fields in the presence of a sphere. In this section we define an alternative basis formed by {\em normalized scattering eigenmodes}. As shown below, these normalized eigenmodes are constructed using the decomposition of a plane wave in spherical waves and, physically, are related to the Lorenz-Mie solutions \cite{Lorenz1890,Mie1908}. The normalized scattering eigenmodes are thus a natural basis for scattering problems. We denote these normalized eigenmodes $\mathbf{F}_\kappa(\rr)$, where $\kappa$ is the eigenmode multi-index. 

\subsubsection{Derivation}

The first step toward defining the normalized scattering eigenmodes is to consider the vacuum scenario, namely free space in the absence of the sphere ($\epsilon=1$ and/or $R=0$). In this scenario, two bases of normalized eigenmodes are of special interest for us. On the one hand, the vacuum spherical eigenmodes $\mathbf{S}^0_\alpha(\rr)$ (spherical waves), defined in \secref{sec:sphericalmodes}. On the other hand, the normalized plane-wave eigenmodes, given by
\begin{equation}
\label{eq:plane_wave}
\mathbf{G}_\kappa(\rr)=\mathbf{G}_g(\kk;\rr)\equiv\frac{\exp(\text{i} \kk \cdot \rr)}{\sqrt{(2\pi)^3}} \mathbf{e}_g.
\end{equation}
We have defined the multi-index $\kappa = (g,\kk)$, where $g \in \lbrace 1,2\rbrace$ is a polarization index and $\kk \in \mathbb{R}^3$ the wave vector. To avoid confusion with the mode index $k$ used in the spherical eigenmodes, we explicitly write the modulus of $\mathbf{k}$ as $\abs{\kk}$. The polarization vectors are defined in terms of the standard polar unit vector $\uv_\theta\equiv\uv_\theta(\theta,\phi)$ and azimuthal unit vector $\uv_\phi\equiv\uv_\phi(\theta,\phi)$ in spherical coordinates. They read $\mathbf{e}_1 \equiv \text{i} \mathbf{e}_{\phi}(\theta_k,\phi_k)$ and $\mathbf{e}_2 \equiv \mathbf{e}_{\theta}(\theta_k,\phi_k)$. Since both plane waves and spherical waves are complete bases, they are related by a linear transformation,
\begin{equation}
\label{eq:plane_wave_expansion}
\mathbf{G}_\kappa(\rr) = \sum_{\alpha} d_{\kappa \alpha} \mathbf{S}^0_\alpha(\rr) ,
\end{equation}
where
\begin{equation}
\label{eq:plane_wave_expansion_coefficients}
d_{\kappa \alpha} = \int_{\mathbb{R}^3}\text{d}\rr \mathbf{G}_\kappa(\rr)\cdot\mathbf{S}^{0*}_\alpha(\rr) .
\end{equation} 
This linear transformation will be essential in the definition of the normalized scattering eigenmodes in the presence of the sphere. It is thus necessary to compute the coefficients $d_{\alpha \kappa}$ analytically. To do so, we first use the standard expansion of a scalar plane wave in spherical harmonics and Bessel functions \cite{Jackson1999}, namely
\begin{align}
\frac{\exp(\text{i} \kk \cdot \rr)}{\sqrt{(2\pi)^3}}=&\sqrt{\frac{2}{\pi}}\sum_{lm}\text{i}^l j_l(|\kk|r)Y_l^{m*}(\theta_k,\phi_k)Y_l^m(\theta,\phi).
\end{align}
Second, we express the polarization vectors $\mathbf{e}_g$ in the basis $(\mathbf{e}_r,\mathbf{e}_\theta,\mathbf{e}_\phi)$. Using the vector spherical harmonics we can write
\begin{align}
\label{eq:pol_expansion_1}
\mathbf{e}_1=&-\sqrt{2\pi}[\mathbf{W}_1^1(\theta,\phi)\exp(-\text{i} \phi_k)-\text{c.c.}],\\
\label{eq:pol_expansion_2}
\mathbf{e}_2=&-\sqrt{2\pi}\cos(\theta_k)[\mathbf{W}_1^1(\theta,\phi)\exp(-\text{i} \phi_k)+\text{c.c.}]-\sqrt{4\pi}\sin (\theta_k) \mathbf{W}_1^0(\theta,\phi).
\end{align}
Third, we introduce these expansions, along with the explicit form of $\mathbf{S}^0_\alpha(\rr)$, into the integrand of \eqnref{eq:plane_wave_expansion_coefficients}. In this way, the integrand is expressed as a sum of terms proportional to two vector spherical harmonics and a single spherical harmonic. These terms, when integrated over the unit sphere, have an analytical expression that is explicitly given in Ref.~\cite{Ivers2008}. After carrying out this integration over solid angle, the coefficients $d_{\alpha \kappa}$ are expressed in terms of radial integrals involving two spherical Bessel functions of the same order, which can also be evaluated analytically using the identity
\begin{equation}
\int_0^\infty\text{d}r r^2 j_l(k r)j_l(|\kk| r)= \frac{\pi}{2 k^2}\delta(k-|\kk|).
\end{equation}
Putting everything together, one obtains that
\begin{equation}
\label{eq:d_in_terms_of_c}
d_{\kappa\alpha}=\frac{c_{lmg}^p}{k} \delta(k-|\kk|),
\end{equation}
where
\begin{align}\label{eq:clmcoeffs}
c_{lm1}^\text{TE}& \equiv \text{i}^{l+1} \mathbf{X}_l^{m*}(\theta_k,\phi_k)\cdot \uv_\phi,\\
c_{lm1}^\text{TM}& \equiv \text{i}^{l-1} \mathbf{X}_l^{m*}(\theta_k,\phi_k)\cdot \uv_\theta,\\
c_{lm2}^\text{TE}& \equiv \text{i}^{l} \mathbf{X}_l^{m*}(\theta_k,\phi_k)\cdot \uv_\theta,\\
\label{eq:clmcoeffs1}
c_{lm2}^\text{TM}& \equiv \text{i}^{l} \mathbf{X}_l^{m*}(\theta_k,\phi_k)\cdot \uv_\phi.
\end{align}
Hence, the decomposition of a normalized plane-wave eigenmode in terms of vacuum normalized spherical eigenmodes, \eqnref{eq:plane_wave_expansion}, can be analytically expressed as
\begin{equation}
\label{eq:plane_wave_expansion_explicit}
\mathbf{G}_g(\kk;\rr) =\frac{1}{|\kk|}\sum_{plm}c_{lmg}^p  \mathbf{S}_{lm}^{0p}(|\kk|;\rr).
\end{equation}

Once the transformation between normalized plane-wave eigenmodes and normalized vacuum spherical eigenmodes has been determined, we can define the normalized scattering eigenmodes in the presence of the sphere. We define the normalized scattering eigenmodes as
\begin{equation}
\label{eq:scattering_eigenmodes}
\mathbf{F}_\kappa (\rr) = \mathbf{F}_g(\kk;\rr)  \equiv \sum_\alpha d_{\kappa \alpha} \mathbf{S}_\alpha (\rr)   =\frac{1}{|\kk|}\sum_{plm}c_{lmg}^p  \mathbf{S}_{lm}^{p}(|\kk|;\rr),
\end{equation}
i.e., these modes are constructed by replacing in \eqnref{eq:plane_wave_expansion_explicit} the vacuum normalized spherical eigenmodes $\mathbf{S}_\alpha^0$ by the normalized spherical eigenmodes in the presence of the dielectric sphere, namely $\mathbf{S}_\alpha$. In other words, the decomposition of normalized scattering eigenmodes, $\mathbf{F}_\kappa (\rr)$, in terms of normalized spherical eigenmodes $\mathbf{S}_{lm}^p(k;\mathbf{r})$ has equal coefficients as the decomposition of a plane wave in normalized spherical eigenmodes $\mathbf{S}_{lm}^{0p}(k;\mathbf{r})$ in the absence of the sphere. The usefulness of this newly defined basis of normalized eigenmodes will become evident in the following.

Let us confirm that the scattering eigenmodes defined in \eqnref{eq:scattering_eigenmodes} are indeed normalized eigenmodes. First, note that they fulfill the eigenmode equation as they are a linear superposition of  normalized eigenmodes with the same eigenfrequency. Second, they are orthonormal, namely
\begin{equation}
\int_{\mathbb{R}^3}\text{d}\rr \epsilon(\rr)\mathbf{F}_\kappa^*(\rr)\cdot \mathbf{F}_{\kappa'}(\rr)=\sum_{\alpha}d^*_{\kappa\alpha}d_{\kappa'\alpha} =\delta_{\kappa \kappa'}.
\end{equation}
The second equality can readily be  demonstrated by using \eqnref{eq:plane_wave_expansion_coefficients}, namely 
\begin{align}
\notag
\sum_{\alpha}d^*_{\kappa\alpha}d_{\kappa'\alpha}&=\int_{\mathbb{R}^6}\text{d}\rr\text{d}\rr'\mathbf{G}^*_\kappa(\rr) \cdot[\bar{\boldsymbol{\delta}}(\rr,\rr') \mathbf{G}_{\kappa'}(\rr')]\\
\label{eq:completeness_relation}
&=\int_{\mathbb{R}^3}\text{d}\rr\mathbf{G}^*_\kappa(\rr)\cdot\mathbf{G}_{\kappa'}(\rr)=\delta_{\kappa \kappa'}.
\end{align}
Here we have defined the transverse delta function $\bar{\boldsymbol{\delta}}(\rr,\rr')$, a real matrix whose Cartesian components read
\begin{equation} \label{eq:delta}
\bar{\delta}_{ij}(\rr,\rr')= \uv_i\cdot[\bar{\boldsymbol{\delta}}(\rr,\rr') \uv_j] \equiv \sum_{\alpha}S^{0*}_{\alpha i}(\rr) S_{\alpha j}^0(\rr'),
\end{equation} 
with  $S^{0}_{\alpha i}(\rr)= \mathbf{S}^{0}_{\alpha }(\rr)\cdot \uv_i$.

Let us remark that in analogy to the normalized spherical eigenmodes in \secref{sec:sphericalmodes}, one can define outgoing (incoming) normalized scattering eigenmodes $\mathbf{F}_\kappa^\text{out}(\rr)$ $(\mathbf{F}_\kappa^\text{in}(\rr))$ by choosing outgoing (incoming) normalized spherical eigenmodes in \eqnref{eq:scattering_eigenmodes}. Moreover, the linear transformation relating these two bases of normalized scattering eigenmodes can be analytically derived. To do so, in analogy with the transverse delta function defined in \eqnref{eq:delta}, we define the $\epsilon$-transverse delta function $\bar{\boldsymbol{\delta}}^\epsilon(\rr,\rr')$, another real matrix with
 Cartesian components 
 \begin{equation}\label{eq:deltatransverse}
\bar{\delta}^\epsilon_{ij}(\rr,\rr')= \uv_i\cdot[\bar{\boldsymbol{\delta}}^\epsilon(\rr,\rr') \uv_j]
 \equiv \sum_{\alpha}\epsilon(\rr')S^{*}_{\alpha i}(\rr) S_{\alpha j}(\rr'),
 \end{equation}
 with $S_{\alpha i}(\rr)= \mathbf{S}_{\alpha }(\rr)\cdot \uv_i$. The $\epsilon$-transverse delta function acts as the identity on $\epsilon-$transverse vector fields $\nabla\cdot [\epsilon(\rr)\mathbf{X}(\rr)]=0$, that is
\begin{equation} \label{eq:deltaepsilon}
    \mathbf{X}(\rr) = \int_{\mathds{R}^3}\text{d}\rr' [\bar{\boldsymbol{\delta}}^\epsilon(\rr,\rr')\mathbf{X}(\rr')].
\end{equation}
 
 We remark that the definition \eqnref{eq:deltatransverse} does not depend on which basis of normalized eigenmodes are used in the expansion.
 This fact allows us to derive the linear transformation relating any two bases of normalized eigenmodes.
In particular, by expressing the $\epsilon-$transverse delta function in terms of incoming normalized scattering eigenmodes, one can directly obtain, using \eqnref{eq:deltaepsilon}, the linear transformation
\begin{align}\label{eq:inoutrelation}
    \mathbf{F}_\kappa^\text{out}(\rr)&= \sum_{\kappa' } \mathbf{F}^\text{in}_{\kappa'}(\rr)\int_{\mathbb{R}^3}\text{d}\rr \epsilon(\rr')\mathbf{F}^{\text{in}*}_{\kappa'}(\rr')\cdot \mathbf{F}^\text{out}_{\kappa}(\rr').
\end{align}
This transformation will play an important role in \secref{sec:onephoton}, where we study the scattering of a single-photon plane-wave state.

\subsubsection{Physical insights: relation to Lorenz-Mie solutions and small-particle limit}

The normalized scattering eigenmodes defined in \eqnref{eq:scattering_eigenmodes} coincide with the Lorenz-Mie solutions. This can be explicitly shown by expressing these eigenmodes in the ``Mie form'', i.e., as a combination of a plane wave plus a scattered spherical wave, as follows.

As discussed in \secref{sec:sphericalmodes}, the normalized scattering eigenmodes can be written as the sum of a free part and a scattering part, $\mathbf{S}_\alpha(\rr)  = \mathbf{S}^0_\alpha(\rr)  + \mathbf{S}^\text{sc}_\alpha(\rr)$. By using the relation \eqnref{eq:plane_wave_expansion}, one can express the normalized scattering eigenmodes \eqnref{eq:scattering_eigenmodes} as $\mathbf{F}_\kappa = {\bf G}_\kappa + \sum_\alpha d_{\kappa \alpha } \mathbf{S}^\text{sc}_\alpha$, that is,
\begin{equation}
\label{eq:mie_form}
\mathbf{F}_g  (\kk;\rr) = {\bf G}_g  (\kk;\rr) + \frac{1}{|\kk|}\sum_{plm}c_{lmg}^p  \mathbf{S}_{lm}^{\text{sc}p}(|\kk|;\rr).
\end{equation}
A normalized scattering eigenmode is thus the sum of a plane wave mode in vacuum and a linear combination of the scattered parts of the normalized spherical eigenmodes in the presence of the sphere. The outgoing scattering modes $\mathbf{F}_\kappa^{\rm out}  (\rr)$, correspond to properly normalized solutions to the Lorenz-Mie problem for incoming plane waves of general propagation direction $\kk$ and polarization $g$.

It is insightful to derive an expression for the normalized scattering eigenmodes in the small-particle limit $q,q'\ll 1$. In this limit, as shown in \secref{sec:sphericalmodes}, the scattered part of the electric-dipole-like spherical eigenmodes, namely those with $\alpha=(\text{TM},1,m,k)$, is much larger than the scattered part of any other spherical eigenmode in the far field. Hence in the far field, and given that all the coefficients $c_{lmg}^p$ are of the same order (see Eqs.~\eqref{eq:clmcoeffs}-\eqref{eq:clmcoeffs1}), one can neglect the contribution of higher-order multipolar modes in \eqnref{eq:mie_form}. After summing over the remaining electric dipole modes, and using the small-particle expansion
\begin{equation}
    \sin \varphi_1^\text{TM}\simeq -2q^3 \frac{\epsilon-1}{\epsilon+2},
\end{equation}
we can express the scattering modes in compact form as
\begin{align}\label{eq:Fsmallparticlelimit}
    \mathbf{F}_\kappa (\rr)\simeq \mathbf{G}_\kappa(\rr) + \mu_0\omega_\kappa^2 \mathbf{\bar G}_0(\rr,\omega_\kappa)\cdot \mathbf{p}.
\end{align}
This expression is analytically derived from the definition of the scattering modes by taking the far-field and the small-particle limit.
This expression is written in terms of the free space Green's tensor~\cite{Tai1994}
\begin{equation}
    \mathbf{\bar G}_0(\rr-\rr',\omega)=
    \left(\mathds{1}+\frac{c^2}{\omega^2}\nabla\nabla\right)\frac{\exp (\text{i} \omega |\rr-\rr'|/c)}{4\pi |\rr-\rr'|},
\end{equation}
with $[\nabla\nabla f]_{ij}\equiv \partial_i\partial_j f$, and the effective dipole moment
\begin{equation}\label{eq:definduceddipole}
 \mathbf{p}\equiv\alpha \mathbf{G}_\kappa(\mathbf{0})=3\epsilon_0 V\frac{\epsilon-1}{\epsilon+2}
 \mathbf{G}_\kappa(\mathbf{0}).
\end{equation} 
The electromagnetic fields associated to \eqnref{eq:Fsmallparticlelimit} correspond to the sum of a plane wave plus the fields emitted by a point electric dipole positioned at the center of the sphere ($\rr=0$), oriented parallel to the plane wave polarization, and oscillating at the plane wave frequency $\omega_\kappa$. The dipole moment $\mathbf{p}$ is induced by the incoming wave, as it is proportional to $\mathbf{G}_\kappa(\mathbf{0})$. The proportionality constant $\alpha$ corresponds to the electrostatic polarizability of a sphere \cite{Jackson1999}. Equation \eqref{eq:Fsmallparticlelimit} thus represents the well-known dipolar profile of the field scattered by a sub-wavelength sphere under plane wave illumination.

\subsection{Plane-Wave Modes}
\label{sec:planwaves}

In the previous sections we have focused on the quantized electromagnetic field in the presence of the dielectric medium \eqnref{eq:epsilon_sphere} by considering normalized eigenmodes. 
Here we focus on plane waves, in order to derive the last two ingredients needed to study scattering of quantum states (see \figref{FigureBigPicture}): first, in \secref{sec:planwavesderivation}, we quantize the electromagnetic field in terms of plane waves. This procedure is subtle since plane waves are neither $\epsilon$-transverse nor normalized eigenmodes in the presence of the sphere. Second, in \secref{sec:planewavestrafo} we derive the canonical transformation (Bogoliubov transformation) relating the creation and annihilation operators of plane waves and of normalized eigenmodes. These two last ingredients complete the toolbox for studying scattering in the next section.

\subsubsection{Quantization with plane-wave modes}\label{sec:planwavesderivation}

Let us focus on quantizing the electromagnetic field using plane waves ${\bf G}_\kappa(\rr)$, as defined in \eqnref{eq:plane_wave} with $\kappa = (g,\kk)$. Importantly, plane waves are not $\epsilon$-transverse, i.e.
\begin{equation}
\nabla \cdot [\epsilon (\rr) {\bf G}_\kappa(\rr) ] \neq 0.
\end{equation}
Hence, they cannot be used to expand $\epsilon$-transverse vector fields, such as the vector potential ${\bf A}(\rr,t)$ in the generalized Coulomb gauge. The known solution to this problem~\cite{Glauber1991} is to expand the vector potential in a particular set of $\epsilon-$transverse eigenmodes $\mathbf{\tilde{G}}_\kappa(\rr)$, defined as
\begin{equation}\label{eq:gauge_transformation_result}
\mathbf{\tilde{G}}_\kappa(\rr) \equiv \int_{\mathbb{R}^3}\text{d}\rr'\frac{\epsilon(\rr')}{\epsilon(\rr)} \spare{\bar{\boldsymbol{\delta}}^\epsilon (\rr',\rr)}^T \mathbf{G}_{\kappa}(\rr')= \sum_\alpha \mathbf{A}_{\alpha}(\rr) \int_{\mathbb{R}^3}\text{d}\rr' \epsilon(\rr') \mathbf{A}^{*}_{\alpha }(\rr')  \cdot \mathbf{G}_{\kappa}(\rr').
\end{equation}
Here we have used the $\epsilon$-transverse delta function $\boldsymbol{\delta}^\epsilon(\rr,\rr')$ defined in \eqnref{eq:deltatransverse}, whose transpose reads $[\bar{\boldsymbol{\delta}}^\epsilon (\rr,\rr')^T]_{ij} = \bar{\delta}^\epsilon_{ji} (\rr,\rr')$.
The normalized eigenmodes $\mathbf{\tilde{G}}_\kappa(\rr)$ are related to the plane-wave modes $\mathbf{G}_\kappa$ via a gauge transformation as derived in~\cite{Wubs2012}. Additionally, these normalized eigenmodes are by definition $\epsilon$-transverse,
 \begin{equation}
\nabla \cdot [\epsilon (\rr) {\bf \tilde G}_\kappa(\rr) ] = 0,
\end{equation}
since they are a linear combination of the $\epsilon$-transverse functions $\mathbf{A}_\alpha(\rr)$.

We can now proceed analogously to \secref{sec:normeigenmodes} and expand the transverse conjugate momentum field $\mathbf{\Pi}(\rr,t)$ using the transverse plane waves $\mathbf{G}_\kappa(\rr)$ and the $\epsilon$-transverse vector field $\mathbf{A}(\rr,t)$ using the $\epsilon$-transverse functions $\mathbf{\tilde G}_\kappa(\rr)$. First, we use \eqnref{eq:expand_vector_potential} and \eqnref{eq:coefficients_vector_potential} for the vector potential and \eqnref{eq:expand_conjugate_momentum} and \eqnref{eq:coefficients_conjugate_momentum} for the conjugate momentum to write
\begin{align} 
\mathbf{A}(\rr,t)& = \int_{\mathbb{R}^3} \text{d}\rr' \bar{\boldsymbol{\delta}}^\epsilon(\rr,\rr') \mathbf{A}(\rr',t),\\
\mathbf{\Pi}(\rr,t)& = \int_{\mathbb{R}^3}  \text{d}\rr' \spare{\bar{\boldsymbol{\delta}}^\epsilon(\rr',\rr)}^T \mathbf{\Pi}(\rr',t).
\end{align}
Then, we introduce in the above expressions the identity
\begin{equation}  \label{eq:transversedeltaG}
\bar{\boldsymbol{\delta}}^\epsilon(\rr,\rr')=\sum_\kappa\mathbf{\tilde{G}}^*_\kappa(\rr)\mathbf{G}_\kappa(\rr'),
\end{equation}
that can be readily demonstrated by using \eqnref{eq:gauge_transformation_result} and recalling \eqnref{eq:deltatransverse}. This leads to the following expansion:

\begin{align}
\label{eq:expand_vector_potential_plane_wave}
\mathbf{A}(\rr,t)&=\sum_{\kappa}A_{\kappa}(t)\mathbf{\tilde G}_\kappa(\rr),\\
\label{eq:expand_conjugate_momentum_plane_wave}
\mathbf{\Pi}(\rr,t)&=\sum_{\kappa}\Pi_{\kappa}(t)\mathbf{G}_\kappa(\rr).
\end{align}
The time-dependent coefficients are given by
\begin{align} \label{eq:coefficients_vector_potential_plane_wave}
A_{\kappa}(t)&=\int_{\mathbb{R}^3}\text{d}\rr \mathbf{G}^*_\kappa(\rr)\cdot \mathbf{A}(\rr,t)\\
\label{eq:coefficients_conjugate_momentum_plane_wave}
\Pi_{\kappa}(t)&=\int_{\mathbb{R}^3}\text{d}\rr \mathbf{\tilde G}_\kappa^*(\rr)\cdot \mathbf{\Pi}(\rr,t).
\end{align}
Note that since $\mathbf{A}(\rr,t)$ and $\mathbf{\Pi}(\rr,t)$ are real, then $A^*_\kappa(t)=A_{-\kappa}(t)$ and  $\Pi_\kappa^*(t)=\Pi_{-\kappa}(t)$, where $-\kappa\equiv (g,-\kk)$.  
We emphasize that in \eqnref{eq:expand_vector_potential_plane_wave} we have expanded the $\epsilon$-transverse vector potential $\mathbf{A}(\rr,t)$ in $\epsilon$-transverse functions $\mathbf{\tilde G}_\kappa(\rr)$, whereas in \eqnref{eq:expand_conjugate_momentum_plane_wave}  we have expanded the transverse conjugate momentum $\mathbf{\Pi}(\rr,t)$ in transverse plane-wave modes $\mathbf{G}_\kappa(\rr)$. This is the key technical step to be able to perform a canonical quantization of the electromagnetic field in the presence of dielectric medium.

In analogy to the expansion coefficients in \secref{sec:normeigenmodes}, the coefficients $A_\kappa(t)$ and $\Pi_\kappa(t)$ in Eqs.~\eqref{eq:coefficients_vector_potential_plane_wave} and \eqref{eq:coefficients_conjugate_momentum_plane_wave} are related by the fact that the expanded fields Eqs.~(\ref{eq:expand_vector_potential_plane_wave}) and (\ref{eq:expand_conjugate_momentum_plane_wave}) must satisfy Hamilton's Eqs.~(\ref{eq:hamilton_vector_potential}) and (\ref{eq:hamilton_conjugate_momentum}). In this case, their coupled dynamical equations read
\begin{align} 
\label{eq:hamilton_vector_potential_plane_wave}
\epsilon_0 \partial_t A_\kappa(t)&=\Pi_\kappa(t)-\sum_{\kappa'}v_{\kappa\kappa'}\Pi_{\kappa'}(t),\\
\label{eq:hamilton_conjugate_momentum_plane_wave}
\partial_t \Pi_\kappa (t)&=-\epsilon_0\omega_\kappa^2A_\kappa(t),
\end{align}
where  
\begin{equation}
v_{\kappa \kappa'} \equiv \frac{\epsilon-1}{\epsilon}\int_{V}\text{d}\rr \mathbf{G}^*_\kappa(\rr)\cdot\mathbf{G}_{\kappa'}(\rr), 
\end{equation}
and $V$ denotes the volume of the sphere. The last term in \eqnref{eq:hamilton_vector_potential_plane_wave} describes a sphere-mediated interaction between plane waves with different wave vectors or polarizations, and arises anytime the fields are not expanded in terms of eigenmodes. Note that in the absence of the sphere ($\epsilon = 1$ and/or $R= 0$), one has $v_{\kappa \kappa'}=0$ and hence the second term in \eqnref{eq:hamilton_vector_potential_plane_wave} would vanish. This reflects the fact that, in the absence of the sphere, plane waves are normalized eigenmodes of Maxwell's equations. 

To continue, we define normal variables via 
\begin{align}
A_\kappa(t)& \equiv-\sqrt{\frac{\hbar}{2\epsilon_0 \omega_\kappa}}[a_\kappa(t)-b_\kappa(t)],\\
\Pi_\kappa(t)& \equiv - \text{i} \sqrt{\frac{\epsilon_0 \hbar  \omega_\kappa}{2}}[a_\kappa(t)+b_\kappa(t)].
\end{align}
Also in this case, the normal variables $a_\kappa(t)$ and $b_\kappa(t)$ are not independent, as the conditions $A^*_\kappa(t)=A_{-\kappa}(t)$ and  $\Pi_\kappa^*(t)=\Pi_{-\kappa}(t)$ imply that $b_\kappa(t)=-a^*_{-\kappa}(t)$. The Hamilton's equations Eqs.~(\ref{eq:hamilton_vector_potential_plane_wave}) and (\ref{eq:hamilton_conjugate_momentum_plane_wave}) can then be condensed into equations of motion only for the normal variables $b_\kappa(t)$, which read
\begin{equation}\label{eq:classical_heisenberg_plane_wave}
\partial_t b_\kappa(t)=-\text{i} \omega_{\kappa} b_\kappa(t)
+2\text{i}\sum_{\kappa'}[V_{\kappa\kappa'}b_{\kappa'}(t)-V_{\kappa-\kappa'}b^*_{\kappa'}(t)],
\end{equation}
where $\omega_\kappa = c \abs{\kk}$ and
\begin{equation}
V_{\kappa\kappa'} =V_{\kappa'\kappa}^* \equiv \frac{\sqrt{\omega_\kappa\omega_{\kappa'}}}{4} v_{\kappa\kappa'}.
\end{equation}  
Again, because plane waves are not eigenmodes in the presence of the sphere, in \eqnref{eq:classical_heisenberg_plane_wave} the normal variable associated to a plane-wave does not evolve as an uncoupled harmonic oscillator but it couples to other plane-wave modes. In terms of the above defined normal variables $b_\kappa(t)$, the fields are given by
\begin{align} 
\mathbf{A}(\rr,t)&=\sum_\kappa \sqrt{\frac{\hbar}{2\epsilon_0 \omega_\kappa}}[b_\kappa(t)\mathbf{\tilde{G}}_\kappa(\rr)+\text{c.c.}],\\
\mathbf{\Pi}(\rr,t)&=- \text{i}  \sum_\kappa \sqrt{\frac{\epsilon_0\hbar \omega_\kappa}{2}}[b_\kappa(t)\mathbf{G}_\kappa(\rr)-\text{c.c.}],
\end{align}
Introducing these fields into the classical electromagnetic Hamiltonian~\eqnref{eq:classcial_hamiltonian}, we obtain
\begin{multline} \label{eq:classicalHnormalvariablesG}
H=\hbar\sum_\kappa \frac{ \omega_\kappa}{2}[b_\kappa^*(t)b_\kappa(t)+b_\kappa(t)b^*_\kappa(t)]\\
-\hbar\sum_{\kappa \kappa'}V_{\kappa \kappa'}[b_\kappa^*(t)-b_{-\kappa}(t)][b_{\kappa'}(t)-b^*_{-\kappa'}(t)].
\end{multline}
As opposed to the classical Hamiltonian expressed with normal variables associated to eigenmodes, \eqnref{eq:classicalHnormalvariables}, the above Hamiltonian expressed with plane-wave normal variables is not diagonal.

Canonical quantization can now be performed by promoting normal variables to operators, $b_\kappa \rightarrow \hat b_\kappa$ and  $b_\kappa^* \rightarrow \hat b^\dagger_\kappa$, and imposing the bosonic commutation rules $[\hat{b}_\kappa,\hat{b}_{\kappa'}^\dagger]=\delta_{\kappa\kappa'}$ and $[\hat{b}_\kappa,\hat{b}_{\kappa'}]=[\hat{b}^\dagger_\kappa,\hat{b}_{\kappa'}^\dagger]=0$. The electric and magnetic field operators are given by
\begin{align}
\hat{\mathbf{E}}(\rr)&=\text{i}\sum_{\kappa}\sqrt{\frac{\hbar \omega_\kappa}{2\epsilon_0}}\spare{\frac{\mathbf{G}_\kappa(\rr)}{\epsilon(\rr)}\hat{b}_\kappa-\text{H.c.}},\\
\hat{\mathbf{B}}(\rr)&=\sum_{\kappa}\sqrt{\frac{\hbar}{2\epsilon_0\omega_\kappa}}[\nabla\times\mathbf{ G}_\kappa(\rr)\hat{b}_\kappa+\text{H.c.}],
\end{align}
and the Hamiltonian operator by
\begin{multline}
\label{eq:hamilton_operator_plane_waves}
\hat{H}=\hbar \sum_\kappa \omega_\kappa \pare{ \hat{b}_\kappa^\dagger\hat{b}_\kappa+\frac{1}{2}}
+\hbar\sum_{\kappa\kappa'}\pare{V_{\kappa-\kappa'}\hat{b}^\dagger_\kappa\hat{b}^\dagger_{\kappa'}+V^*_{\kappa-\kappa'}\hat{b}_\kappa \hat{b}_{\kappa'}}\\
-\hbar\sum_{\kappa\kappa'}\left(V_{\kappa\kappa'}\hat{b}^\dagger_\kappa\hat{b}_{\kappa'}+V^*_{\kappa\kappa'}\hat{b}_\kappa\hat{b}^\dagger_{\kappa'}\right).
\end{multline}
Note that this result coincides with the Hamiltonian derived in \cite{Glauber1991}. The above Hamiltonian can be expressed in a more familiar way using the electric field operator, namely

\begin{equation} \label{eq:hamilton_operator_plane_waves_intuitive}
\hat{H}=\hbar \sum_\kappa \omega_\kappa \pare{ \hat{b}_\kappa^\dagger\hat{b}_\kappa+\frac{1}{2}}
-\frac{\epsilon_0}{2}\epsilon(\epsilon-1) \int_{V} \text{d}\rr \hat{\mathbf{E}}^2(\rr).
\end{equation}
The total Hamiltonian is expressed as the sum of the plane-wave Hamiltonian in free space plus an extra term that accounts for the presence of the dielectric sphere and is responsible for the coupling between different plane waves.
We remark that this second term differs from the expression commonly used and heuristically derived in the literature~\cite{Chang2010,  Barker2010, Romero-Isart2010, Romero-Isart2011, Pflanzer2012, Tebbenjohanns2019, GonzalezBallestero2019, Toros2020} for a \textit{moving} levitated nanoparticle in the quantum regime, which is explicitly proportional to the particle polarizability $\alpha$ (\eqnref{eq:definduceddipole}). The conclusions that follow from this discrepancy will be further discussed in a subsequent article, where the motion of the sphere is included as a dynamical variable.

\subsubsection{Canonical Transformation between Normalized Eigenmodes and Plane-Wave Modes}
\label{sec:planewavestrafo}

So far we have shown how the Hamiltonian of the system can be derived both in terms of normalized eigenmodes $\mathbf{A}_\alpha(\rr)$ with the operators $\hat{a}_\alpha,\hat{a}_\alpha^\dagger$, see \eqnref{eq:hamilton_operator_eigenmodes}, and plane-wave modes $\mathbf{G}_\kappa(\rr)$ with the operators $\hat{b}_\kappa,\hat{b}_\kappa^\dagger$, see \eqnref{eq:hamilton_operator_plane_waves}. Here we derive the canonical transformation relating these two sets of bosonic operators. This is the last crucial ingredient to study quantum scattering in \secref{sec:qmie}.

We start by establishing a relation between the coefficients $A_\alpha(t)$ and $A_\kappa(t)$, defined by \eqnref{eq:expand_vector_potential} and \eqnref{eq:expand_vector_potential_plane_wave}, respectively. This can be done by inserting Eqs.~(\ref{eq:gauge_transformation_result}) and (\ref{eq:expand_vector_potential_plane_wave}) into \eqnref{eq:coefficients_vector_potential} to obtain \begin{align}
A_\alpha(t)&=\sum_\kappa A_\kappa (t)\int_{\mathbb{R}^3}\text{d}\rr \epsilon(\rr)\mathbf{A}^*_\alpha(\rr)\cdot \mathbf{G}_\kappa(\rr).
\end{align}
The inverted relation is obtained by inserting \eqnref{eq:expand_vector_potential} into  \eqnref{eq:coefficients_vector_potential_plane_wave}, which leads to
\begin{equation}
A_\kappa(t)=\sum_\alpha A_\alpha (t)\int_{\mathbb{R}^3}\text{d}\rr \mathbf{G}^*_\kappa(\rr)\cdot \mathbf{A}_\alpha(\rr).
\end{equation}
Similar relations can be derived for the time-dependent coefficients of the conjugate momentum field. Specifically, inserting \eqnref{eq:expand_conjugate_momentum_plane_wave} into \eqnref{eq:coefficients_conjugate_momentum} leads to
\begin{align}
\Pi_\alpha(t)&=\sum_\kappa \Pi_\kappa (t)\int_{\mathbb{R}^3}\text{d}\rr \mathbf{A}^*_\alpha(\rr)\cdot \mathbf{G}_\kappa(\rr),
\end{align}
while inserting \eqnref{eq:expand_conjugate_momentum} into \eqnref{eq:coefficients_conjugate_momentum_plane_wave} leads to
\begin{align}
\Pi_\kappa(t)&=\sum_\alpha \Pi_\alpha (t)\int_{\mathbb{R}^3}\text{d}\rr \epsilon(\rr) \mathbf{G}^*_\kappa(\rr)\cdot\mathbf{A}_\alpha(\rr).
\end{align}

We now carry out the same steps as for the canonical quantization, namely we perform a change of variables to normal variables, and then replace such normal variables by operators. This transforms the above expressions directly into a relation between the corresponding operators,
\begin{align}
\label{eq:bkappa_in_terms_of_aalpha}
\hat{b}_\kappa&=\sum_\alpha A_{\kappa\alpha}\hat{a}_\alpha+B_{\kappa\alpha}\hat{a}^\dagger_\alpha,\\
\label{eq:aalpha_in_terms_of_bkappa}
\hat{a}_\alpha&=\sum_\kappa A^*_{\kappa\alpha}\hat{b}_\kappa-B_{\kappa\alpha}\hat{b}^\dagger_\kappa.
\end{align}
The coefficients defining the transformation are given by 
\begin{align}
\label{eq:Akappaalpha}
A_{\kappa\alpha}&\equiv\sqrt{\frac{\omega_\alpha}{\omega_\kappa}}\int_{\mathbb{R}^3}\text{d}\rr \left[\frac{\omega_\kappa}{2\omega_\alpha}+\frac{\epsilon(\rr)}{2}\right]\mathbf{G}^*_\kappa(\rr)\cdot \mathbf{A}_{\alpha}(\rr),\\
\label{eq:Bkappaalpha}
B_{\kappa\alpha}&\equiv\sqrt{\frac{\omega_\alpha}{\omega_\kappa}}\int_{\mathbb{R}^3}\text{d}\rr \left[\frac{\omega_\kappa}{2\omega_\alpha}-\frac{\epsilon(\rr)}{2}\right]\mathbf{G}^*_\kappa(\rr)\cdot \mathbf{A}^*_{\alpha}(\rr).
\end{align}
The fact that this is a canonical transformation follows from the relations
\begin{align}
\sum_{\alpha}A_{\kappa \alpha}A^{*}_{\kappa' \alpha}-B_{\kappa \alpha}B^{*}_{\kappa' \alpha}&=\delta_{\kappa\kappa'},\\
\sum_{\kappa}A^{*}_{\kappa \alpha}A_{\kappa \alpha'}-B_{\kappa \alpha}B^{*}_{\kappa \alpha'}&=\delta_{\alpha\alpha'},\\
\sum_{\alpha}A_{\kappa \alpha}B_{\kappa' \alpha}-B_{\kappa \alpha}A_{\kappa' \alpha}&=0,\\
\sum_{\kappa} A^{*}_{\kappa \alpha}B_{\kappa \alpha'}- B_{\kappa \alpha}A^{*}_{\kappa \alpha'}&=0,
\end{align}
which can be explicitly verified using the properties of both the normalized eigenmodes and the plane-wave modes. 

The canonical transformation introduced here can also be understood as a Bogoliubov transformation that diagonalizes the Hamiltonian \eqnref{eq:hamilton_operator_plane_waves} into \eqnref{eq:hamilton_operator_eigenmodes}. As a side remark, note that by expressing $\mathbf{A}(\rr,t)$ in two arbitrary sets of normalized eigenmodes, $\mathbf{A}_{1\alpha}(\rr)$ and $\mathbf{A}_{2\beta}(\rr)$, one can  readily derive a linear transformation for the corresponding operators following analogous steps, namely
\begin{equation}\label{eq:lineartrafoem}
    \hat{a}_{1\alpha}= \sum_\beta \hat{a}_{2\beta} \int_{\mathds{R}^3}\text{d}\rr \epsilon(\rr) \mathbf{A}_{1\alpha}^*(\rr)\cdot \mathbf{A}_{2\beta}(\rr).
\end{equation}

In \secref{sec:onephoton} we will make use of the above canonical transformation for the particular case of normalized scattering eigenmodes. It is thus convenient to simplify Eqs.~(\ref{eq:Akappaalpha}) and (\ref{eq:Bkappaalpha}) for the case of normalized scattering eigenmodes, i.e. for the particular case $\mathbf{A}_\alpha(\mathbf{r})\to\mathbf{F}_{\kappa'}(\rr)$. First, note that we can write
\begin{align}
A_{\kappa\kappa'}&=\sqrt{\frac{\omega_{\kappa'}}{\omega_\kappa}}\left(\frac{\omega_\kappa \chi_{\kappa\kappa'}^0}{2\omega_{\kappa'}}+\frac{ \chi_{\kappa\kappa'}^1}{2}\right),\\
B_{\kappa\kappa'}&=\sqrt{\frac{\omega_{\kappa'}}{\omega_\kappa}}\sum_{\kappa''}M_{\kappa'\kappa''}\left(\frac{\omega_\kappa \chi_{\kappa\kappa''}^0}{2\omega_{\kappa'}}-\frac{ \chi_{\kappa\kappa''}^1}{2}\right),
\end{align}
where we have defined
\begin{equation}
\label{eq:chi_nu}
\chi^\nu_{\kappa\kappa'}\equiv \int_{\mathbb{R}^3}\text{d}\rr \spare{\epsilon(\rr)}^\nu \mathbf{G}^*_\kappa(\rr)\cdot \mathbf{F}_{\kappa'}(\rr),
\end{equation}
with $\nu \in \lbrace 0,1 \rbrace$, and used the matrix $M_{\kappa \kappa'}$ defined in \eqnref{eq:Mmatrix}. In order to evaluate \eqnref{eq:chi_nu} we express both the plane-wave mode and the normalized scattering eigenmode in terms of the normalized spherical eigenmodes using Eqs.~(\ref{eq:plane_wave_expansion}) and (\ref{eq:scattering_eigenmodes}). The angular integral can be evaluated using the orthonormality relations of the vector spherical harmonics, whereas the radial integrals can be evaluated using the techniques reported in~\cite{McPhedran2020}. This leads to
\begin{multline}\label{eq:Akk_scattering_modes}
A_{\kappa\kappa'}= \sum_{plm}c_{lmg}^{p*}c_{lmg'}^{p}\exp(\mp \text{i} \varphi_l^p)\cos\varphi_l^p\frac{\delta(|\kk|-|\kk'|)}{|\kk|^2}\\
+\frac{\epsilon-1}{2}\frac{\sqrt{|\kk||\kk'|}}{|\kk|-|\kk'|}\int_V\text{d}\rr \mathbf{G}_\kappa^*(\rr)\cdot \mathbf{F}_{\kappa'}(\rr)
\end{multline}
and
\begin{equation}
\label{eq:Bkk_scattering_modes}
B_{\kappa\kappa'}=-\frac{\epsilon-1}{2}\frac{\sqrt{|\kk||\kk'|}}{|\kk|+|\kk'|}\int_V\text{d}\rr \mathbf{G}_\kappa^*(\rr)\cdot \mathbf{F}^*_{\kappa'}(\rr).
\end{equation}
Note that the result depends on whether one considers outgoing $(-\text{i}\varphi_l^p)$ or incoming $(+\text{i}\varphi_l^p)$ normalized scattering eigenmodes. We can now insert Eqs.~(\ref{eq:Akk_scattering_modes}) and (\ref{eq:Bkk_scattering_modes}) into \eqnref{eq:bkappa_in_terms_of_aalpha} to obtain an explicit expression for the canonical transformation. Since the second term in \eqnref{eq:Akk_scattering_modes} has a pole we employ the Sokhotski–Plemelj theorem in order to perform the integration along $|\kk'|\geq 0$. We finally obtain the following explicit expression:
\begin{multline} \label{eq:explicitAkkBkk}
\hat{b}_\kappa=\hat{a}_\kappa
+\sum_{\kappa'} \frac{\epsilon-1}{2}\sqrt{|\kk||\kk'|}\int_V\text{d}\rr\frac{\mathbf{G}^*_\kappa(\rr)\cdot \mathbf{F}_{\kappa'}(\rr)}{|\kk|-|\kk'|\mp\text{i} \eta}\hat{a}_{\kappa'}\\
-\sum_{\kappa'}\frac{\epsilon-1}{2}\sqrt{|\kk||\kk'|}\int_V\text{d}\rr\frac{\mathbf{G}^*_\kappa(\rr)\cdot \mathbf{F}^*_{\kappa'}(\rr)}{|\kk|+|\kk'|}\hat{a}^\dagger_{\kappa'},
\end{multline}
for outgoing ($-\text{i} \eta$) and incoming ($+\text{i} \eta$) normalized scattering eigenmodes. Note that \eqnref{eq:explicitAkkBkk} expresses the creation and annihilation operators of a photon in a plane-wave mode in terms of the creation and annihilation operators of an either incoming or outgoing photon in a scattering eigenmode. These relations, along with all the theoretical framework derived in \secref{sec:quantization}, can now be employed to describe quantum Lorenz-Mie scattering.

\section{Quantum Lorenz-Mie Scattering}
\label{sec:qmie}

In this section we use the theoretical formalism developed in \secref{sec:quantization} to study the scattering of quantum states of light off a dielectric sphere. We focus on four particular problems of relevance to optical levitodynamics. First, in \secref{sec:onephoton} we study the scattering of a single-photon plane-wave state, and exactly derive the single-photon scattering matrix, the transition amplitude, and both the differential and the total scattering cross sections. From these quantities, the scattering of any single-photon pulse can readily be obtained. Second, in \secref{sec:cohsqueezed}, we consider a sphere under continuous illumination from a coherent and potentially squeezed light source, which models the optical tweezers used in optical control and readout of levitated particles. We calculate the power scattered by the sphere and the corresponding scattering cross section. Third, in \secref{sec:twophoton}, we study the scattering of a two-photon state off the dielectric sphere as an illustration of quantum effects whose description requires a quantum theoretical formalism. Specifically, by calculating the angular distribution of two-photon correlations in the far-field, we unveil quantum interference effects due to Hong-Ou-Mandel effect. We emphasize that since the sphere is nonmoving and rigid, the considered scattering processes turn out to be elastic, i.e. the photons do not change their frequency. Finally, we conclude our discussion of two-photon scattering in \secref{sec:2p_transition_amplitude} by deriving the two-photon scattering amplitude and cross section for a two-photon plane wave state.

\subsection{Single-Photon State}
\label{sec:onephoton}

Here we study the scattering of a single-photon plane wave state. For this purpose, we aim at calculating the transition amplitude from the initial state $\hat{b}^\dagger_{\kappa'}\ket{0}$ at time $t'$ to a final state $\hat{b}^\dagger_{\kappa}\ket{0}$ at time $t>t'$, where $\hat{b}^\dagger_{\kappa}$ is the creation operator associated to the plane-wave mode $\mathbf{G}_\kappa (\rr) = \mathbf{G}_g (\kk;\rr)$, as defined in \secref{sec:planwaves}, and $\ket{0}$ is the vacuum state \textit{in the presence of the dielectric sphere}. This transition amplitude can be written as
\begin{equation}
\label{eq:transition_amplitude}
\mathcal{T}_{\kappa\kappa'} (t,t') = \bra{0}\hat{b}_{\kappa}\hat{U}(t,t')\hat{b}^\dagger_{\kappa'}\ket{0}= \bra{0}\hat{b}_{\kappa}(t)\hat{b}^\dagger_{\kappa'}(t')\ket{0}.
\end{equation}
The unitary time-evolution operator $\hat{U}(t,t')$  fulfills the Schr\"odinger equation $\text{i}\hbar\partial_t \hat{U}(t,t')=\hat{H}\hat{U}(t,t')$ with $\hat{U}(t,t)=\mathds{1}$. In the right-hand side of Eq.~\eqref{eq:transition_amplitude} we have used the property $\hat{U}(t,t')=\hat{U}(t,0)\hat{U}^\dagger(t',0)$, the fact that the vacuum state is an eigenstate of $\hat{H}$ -- that is, $\ket{0}= \hat U(t,t') \ket{0}$ -- and we have written the photonic creation and annihilation operators in the Heisenberg picture via $\hat{b}_\kappa(t) = \hat{U}^\dagger(t,0) \hat{b}_\kappa \hat{U}(t,0)$.

We are interested in time-independent scattering parameters (e.g. scattering cross section) that account for the whole scattering process~\cite{Newton1982, Goldberger2004, Taylor2006}. We thus focus on 
calculating the asymptotic transition amplitude given by $\mathcal{T}_{\kappa\kappa'} \equiv \mathcal{T}_{\kappa\kappa'} (t_+,t_-) $, where hereafter we use the shorthand notation $f(t_{\pm}) = \lim_{t\rightarrow \pm \infty} f(t)$. To compute this amplitude, we use an adiabatic approximation, i.e. we consider a slowly varying, time-dependent relative permittivity
\begin{equation}
\epsilon(\rr,t)=1+(\epsilon-1)\exp(-\eta|t|)\Theta(R-|\rr|),
\end{equation}
where the parameter $\eta>0$ is made arbitrarily small \cite{Glauber1991}. Effectively, this describes the interaction of the scatterer with a quasi-monochromatic single-photon pulse. In the limit of small $\eta$ one can approximate $\partial_t[\epsilon(\rr,t)\mathbf{E}(\rr,t)]\simeq \epsilon(\rr,t)\partial_t\mathbf{E}(\rr,t)$, which allows us to define the instantaneous normalized eigenmodes fulfilling
\begin{equation}
\nabla\times\nabla\times\mathbf{A}_\alpha(\rr,t)-\epsilon(\rr,t)\frac{\omega_\alpha^2}{c^2}\mathbf{A}_\alpha(\rr,t)=0.
\end{equation}
They are easily constructed from the normalized eigenmodes $\mathbf{A}_\alpha(\rr)=\mathbf{A}_\alpha(\rr,0)$, which are solutions of \eqnref{eq:eigenmode_equation}, by replacing $\epsilon \rightarrow1+(\epsilon-1)\exp(-\eta |t|)$. 

Within the adiabatic approximation, the single-photon scattering matrix can be exactly calculated as follows. From \eqnref{eq:hamilton_operator_plane_waves_intuitive}, one can show that the Heisenberg equation for the operator $\tilde{b}_\kappa(t)\equiv \hat{b}_\kappa(t) \exp(\text{i} \omega_\kappa t)$ is given by
\begin{equation}
\partial_t \tilde{b}_\kappa(t)=(\epsilon-1)\sqrt{\frac{\epsilon_0\omega_\kappa}{2\hbar}}
\label{eq:bkappa_heisenberg_equation}
\int_V\text{d}\rr \mathbf{G}^*_\kappa(\rr)\cdot \mathbf{E}(\rr,t)\exp(\text{i} \omega_\kappa t-\eta |t|).
\end{equation}
We now express $\mathbf{E}(\rr,t)$ in terms of the instantaneous normalized scattering eigenmodes $\mathbf{F}_\kappa(\rr,t)$ without yet specifying their far-field behavior, and formally integrate \eqnref{eq:bkappa_heisenberg_equation}. In this step, the time integral on the right-hand side can be approximated using integration by parts as
\begin{align}
\int_{t_\pm}^0\text{d}t \exp[(\text{i} \omega\mp \eta)t]\mathbf{F}_\kappa(\rr,t)\simeq \frac{\mathbf{F}_\kappa(\rr)}{\text{i} \omega \mp \eta},
\end{align}
up to corrections of order $\eta$ which, within the adiabatic approximation, can be made arbitrarily small.
The result for the plane wave operators is
\begin{multline}\label{eq:solution_heisenberg_equation}
\hat{b}_\kappa = \tilde{b}_\kappa(t_\mp)+\frac{\epsilon-1}{2}\sum_{\kappa'}\sqrt{|\kk||\kk'|}\int_V\text{d}\rr \frac{\mathbf{G}_\kappa^*(\rr)\cdot \mathbf{F}_{\kappa'}(\rr)}{|\kk|-|\kk'|\mp \text{i} \eta}\hat{a}_{\kappa'}\\
-\frac{\epsilon-1}{2}\sum_{\kappa'}\sqrt{|\kk||\kk'|}\int_V\text{d}\rr \frac{\mathbf{G}_\kappa^*(\rr)\cdot \mathbf{F}^*_{\kappa'}(\rr)}{|\kk|+|\kk'|}\hat{a}^\dagger_{\kappa'}.
\end{multline}
The key step is now to compare \eqnref{eq:explicitAkkBkk} with \eqnref{eq:solution_heisenberg_equation}, which allows us to conclude that,
\begin{align}
\label{eq:id1}
\tilde{b}_\kappa (t_-) &= \hat{a}_\kappa^\text{out},\\
\label{eq:id2}
\tilde{b}_\kappa (t_+) &= \hat{a}_\kappa^\text{in}.
\end{align}
That is, the asymptotic operator $\tilde{b}_\kappa (t_-)$ ($\tilde{b}_\kappa (t_+)$) is equal to the outgoing scattering eigenmode operator $\hat{a}_\kappa^\text{out}$ (incoming scattering eigenmode operator $\hat{a}_\kappa^\text{in}$). Recall that the incoming/outgoing normalized scattering eigenmodes are linearly dependent, see \eqnref{eq:inoutrelation}. As a consequence of \eqnref{eq:lineartrafoem}, the corresponding scattering eigenmode operators are related by the same linear transformation. This transformation defines the single-photon scattering matrix~\cite{Taylor2006}, 
\begin{equation}\label{eq:asymptotic_transformation}
    \tilde{b}_\kappa(t_+)=\sum_{\kappa'}S_{\kappa \kappa'}\tilde{b}_{\kappa'}(t_-),
\end{equation}
and reads
\begin{equation}
S_{\kappa\kappa'}=\int_{\mathbb{R}^3}\text{d}\rr \epsilon(\rr)\mathbf{F}^{\text{in}*}_\kappa(\rr)\cdot \mathbf{F}^\text{out}_{\kappa'}(\rr)
=\sum_{plm}c_{lmg}^{p*}c_{lmg'}^{p}\exp(-2\text{i} \varphi_l^p)\frac{\delta(|\kk|-|\kk'|)}{|\kk|^2}.
\end{equation}
The second equality is obtained by inserting \eqnref{eq:scattering_eigenmodes} and using the orthonormality of the normalized spherical eigenmodes. 

Once the scattering matrix is obtained, the transition amplitude can be readily computed. Using the relation $\exp(- 2 \text{i} x )=1-2\text{i} \sin(x)\exp(-\text{i} x)$, \eqnref{eq:d_in_terms_of_c} and \eqnref{eq:completeness_relation} we can express the scattering matrix as
\begin{equation}
\label{eq:scatt_mat}
S_{\kappa\kappa'}=\delta_{\kappa\kappa'}+\frac{\text{i} f_{\kappa\kappa'}}{2\pi |\kk|}\delta(|\kk|-|\kk'|),
\end{equation}
where the scattering amplitude is defined as
\begin{equation}
\label{eq:scattering_amplitude_elastic}
f_{\kappa \kappa'}=-\frac{4\pi}{|\kk|}\sum_{plm}c_{lmg}^{p*}c_{lmg'}^{p}\sin(\varphi_l^p)\exp(-\text{i} \varphi_l^p).
\end{equation}
From these results one obtains that the asymptotic transition amplitude for single-photon states reads
\begin{equation}
\label{eq:transition_amplitude_elastic}
\mathcal{T}_{\kappa \kappa'}=\left[\delta_{\kappa\kappa'}+\frac{\text{i} f_{\kappa \kappa'}}{2 \pi |\kk|}\delta(|\kk|-|\kk'|)\right]\exp[-\text{i} \omega_\kappa(t_+-t_-)].
\end{equation}

We can finally derive the single-photon scattering cross section for our initial and final states, both of which belong to the same continuous spectrum. By analogy to the transition probability between two states of a discrete spectrum, one can derive a scattering cross section quantifying the transition probability per unit time to all possible final states in units of an incident flux~\cite{Cohen2004/2}
\begin{align}
\label{eq:definition_differential_scattering_cs}
\sigma&\equiv  \int_{\mathbb{S}^2}\text{d}\Omega_k\frac{\text{d}\sigma}{\text{d}\Omega_{\kappa}}=\int_{\mathbb{S}^2}\text{d}\Omega_k  \sum_g|f_{\kappa\kappa'}|^2,
\end{align}
where $\text{d}\sigma/\text{d}\Omega_{\kappa}$ denotes the differential scattering cross section, i.e. the scattering cross section per unit solid angle. Inserting the scattering amplitude \eqnref{eq:scattering_amplitude_elastic} we obtain
\begin{align}
\label{eq:single_photon_scattering_cross_section}
\sigma&=\frac{16 \pi^2}{|\kk|^2}\sum_{plm}|c_{lmg}^p\sin\varphi_l^p|^2\\
&=\frac{2\pi}{|\kk|^2}\sum_{lp}(2l+1)\sin^2(\varphi_l^p),
\end{align}
where the first line follows from \eqnref{eq:completeness_relation} together with \eqnref{eq:d_in_terms_of_c} and the second line follows from the spherical harmonics addition theorem
\begin{equation}
\label{eq:spherical_addition}
\sum_{m}Y_{l}^{m*}(\theta',\phi')Y_{l}^m(\theta,\phi) = \frac{2l+1}{4\pi}P_l(
\cos\gamma),
\end{equation}
where $P_l(x)$ denotes the Legendre polynomial of degree $l$ and $\cos\gamma= \cos\theta\cos\theta'+\sin\theta\sin\theta'\cos(\phi-\phi')$. Finally, note that the optical theorem is fulfilled as
the scattering amplitude in the forward direction can be expressed in terms of the scattering cross section via
\begin{equation}
\text{Im}[f_{\kappa\kappa}]=\frac{k\sigma}{4\pi}.
\end{equation}

As an example, let us particularize to an initial plane-wave single-photon state polarized along the $x-$axis and traveling along the $z-$axis. The differential and total scattering cross sections take a particularly simple form in the small-particle limit $q\ll 1$, where they read 
\begin{align}
\frac{\text{d}\sigma}{\text{d}\Omega_k}&=\frac{3\sigma}{8\pi}[1-(\uv_k\cdot \uv_x)^2],\\
\label{eq:scattering_cross_section_elastic_small}
\sigma&=\frac{8\pi}{3}\left(\frac{\epsilon-1}{\epsilon+2}q^2R\right)^2.
\end{align}
These expressions are equal to the Rayleigh differential and total scattering cross sections \cite{Bohren2004}. More generally, as we will show in \secref{sec:cohsqueezed}, the single-photon scattering cross section is equal to the classical Lorenz-Mie scattering cross section up to all orders in $q$.

\subsection{Coherent Squeezed State}
\label{sec:cohsqueezed}

We now focus on the scattering of a continuous light beam off the dielectric sphere, specifically of a coherent and squeezed beam. We thus consider that the state of the electromagnetic field
in the presence of the dielectric sphere is described by the following coherent squeezed state,
\begin{equation} \label{initialsqueezedstate}
\ket{\alpha_\kappa,\zeta_{\kappa}}= \hat D(\alpha_{\kappa}) \hat S(\zeta_{\kappa})\ket{0}.
\end{equation}
Here, the displacement and the squeezing operator (in the single-mode-product form, without considering entanglement between modes) are given, respectively, by
\begin{align}
\label{eq:displacement_operator}
\hat{D}(\alpha_\kappa)&=\exp \left[\sum_\kappa (\alpha_{\kappa}\hat{a}_\kappa^\dagger-\alpha^*_\kappa \hat{a}_\kappa)\right],\\
\label{eq:squeezing_operator}
\hat{S}(\zeta_{\kappa})&=\exp \left[\sum_{\kappa} \left(\frac{\zeta^*_{\kappa}}{2}\hat{a}_\kappa\hat{a}_\kappa-\frac{\zeta_{\kappa}}{2}\hat{a}^\dagger_\kappa\hat{a}^\dagger_\kappa\right)\right],
\end{align}
where $\hat{a}_\kappa$ and $\hat{a}^\dagger_\kappa$ are the creation and annihilation operators associated to a \textit{normalized scattering eigenmode} $\mathbf{F}_\kappa(\rr)$, as introduced in \secref{sec:scatteringmodes}.
In order to model an incoming optical beam, we consider the outgoing eigenmodes, but do not specify this explicitly in the notation for the sake of compactness. 
 The vacuum state $\ket{0}$ in the presence of the sphere is defined by $\hat{a}_\kappa \ket{0}=0$. The displacement and squeezing of the mode $\kappa$ is given by $\alpha_\kappa = \abs{\alpha_\kappa} \exp(\text{i} \theta_\kappa) \in \mathbb{C}$ and $\zeta_{\kappa}= r_\kappa\exp(\text{i} \phi_\kappa)$, respectively. Phase-squeezed light corresponds to $\theta_\kappa = (\phi_\kappa+\pi)/2$ and amplitude-squeezed light to $\theta_\kappa = \phi_\kappa/2$.  Note that the state \eqnref{initialsqueezedstate} is not an eigenstate of the Hamiltonian.

For simplicity we assume that only a given mode, denoted by the multi-index $\kappa_0$, is displaced and squeezed, so that $\alpha_\kappa = \alpha \delta_{\kappa \kappa_0}$  and, for arithmetic convenience, $r_\kappa = \text{arcsinh}(\rho\sqrt{ \delta_{\kappa\kappa_0}})$ with $\rho\in\mathbb{R}$, which can be approximated by any sufficiently peaked function at $\kappa_0$. We then evaluate the time-averaged expectation value, over a single period of the mode $\kappa_0$, of the Poynting vector operator,
\begin{equation}\label{eq:Poyntingvectordef}
\mathbf{P}(\rr)\equiv\langle\langle\hat{\mathbf{P}}(\rr,t)\rangle\rangle_T  = \frac{1}{T} \int_{-T/2}^{T/2} \text{d}t  \bra{\alpha_\kappa,\zeta_{\kappa}} \hat{\mathbf{P}}(\rr,t) \ket{\alpha_\kappa,\zeta_{\kappa}},
\end{equation}
which in the Heisenberg picture reads
\begin{equation}
\hat{\mathbf{P}}(\rr,t)=\frac{\hat{\mathbf{E}}(\rr,t)\times \hat{\mathbf{B}}(\rr,t)-\hat{\mathbf{B}}(\rr,t)\times \hat{\mathbf{E}}(\rr,t)}{2\mu_0}.
\end{equation}
To perform the calculation we expand the electric and magnetic field operators in normalized scattering eigenmodes (recall \eqnref{eq:electric_operator_eigenmodes} and \eqnref{eq:magnetic_operator_eigenmodes}), namely
\begin{align}
\hat{\mathbf{E}}(\rr,t)&=\text{i}\sum_{\kappa}\sqrt{\frac{\hbar \omega_\kappa}{2\epsilon_0}}\spare{\mathbf{F}_\kappa(\rr)\hat{a}_\kappa e^{\text{i} \omega_\kappa t}-\text{H.c.}},\\
\hat{\mathbf{B}}(\rr,t)&=\sum_{\kappa}\sqrt{\frac{\hbar}{2\epsilon_0\omega_\kappa}}\spare{\nabla\times\mathbf{F}_\kappa(\rr)\hat{a}_\kappa e^{\text{i} \omega_\kappa t} +\text{H.c.}}.
\end{align}
 Then, we derive the following relations using Hadamard's lemma,
\begin{equation}
\label{eq:hadamard1}
\hat{D}^\dagger(\alpha_\kappa)\hat{a}_{\kappa'}\hat{D}(\alpha_\kappa) = \hat{a}_{\kappa'}+\alpha_{\kappa'},
\end{equation}
\begin{equation}
\label{eq:hadamard2}
\hat{S}^\dagger(\zeta_{\kappa})\hat{a}_{\kappa'}\hat{S}(\zeta_{\kappa}) = \cosh(r_{\kappa'})\hat{a}_{\kappa'}
-\sinh(r_{\kappa'})\exp(\text{i} \phi_{\kappa'})\hat{a}^\dagger_{\kappa'},
\end{equation}
and use them to compute the averages
\begin{align}
\langle\langle \hat{a}_\kappa(t)\hat{a}_{\kappa'}(t)\rangle\rangle_T&=0,\\
\langle\langle \hat{a}_\kappa(t)\hat{a}^\dagger_{\kappa'}(t)\rangle\rangle_T&=\alpha_\kappa\alpha^*_{\kappa'}+\cosh^2(r_\kappa)\delta_{\kappa\kappa'},\\
\langle\langle \hat{a}^\dagger_\kappa(t)\hat{a}_{\kappa'}(t)\rangle\rangle_T&=\alpha^*_\kappa\alpha_{\kappa'}+\sinh^2(r_\kappa)\delta_{\kappa\kappa'},\\
\langle\langle \hat{a}^\dagger_\kappa(t)\hat{a}^\dagger_{\kappa'}(t)\rangle\rangle_T&=0.
\end{align}

Following the above steps, the resulting Poynting vector \eqnref{eq:Poyntingvectordef} can be split as
\begin{equation}\label{eq:Pvectordecomposition}
\mathbf{P}(\rr)= \mathbf{P}_\text{vac}(\rr) + \mathbf{P}_\text{cl}(\rr)+  \mathbf{P}_\text{sq}(\rr).
\end{equation}
The first term describes the vacuum contribution, i.e. the Poynting vector associated to the vacuum state of the electromagnetic field $\vert 0 \rangle$ (or, equivalently, to the case $\alpha = \rho = 0$). It is independent on $\kappa_0$ and given by
\begin{equation}\label{eq:Svac}
    \frac{\mathbf{P}_\text{vac}(\rr)}{\hbar c^2} =
    -\frac{1}{2}\sum_\kappa \text{Im}[\mathbf{F}_{\kappa}(\rr)\times \nabla\times \mathbf{F}_{\kappa}^*(\rr)].
\end{equation}
Analogous to the zero-point energy, this purely quantum contribution to the scattered power can only be obtained using a quantum theory. In the far field, the radiated power associated to this Poynting vector vanishes
\begin{equation}
    \lim_{|\rr|\rightarrow\infty} |\rr|^2 \int_{\mathbb{S}^2}\text{d}\Omega \mathbf{P}_\text{vac}(\rr)=0,
\end{equation}
as can be shown inserting \eqnref{eq:scattering_eigenmodes} in \eqnref{eq:Svac} and using the far-field expansion of the normalized spherical eigenmodes. The second and third contributions to \eqnref{eq:Pvectordecomposition} stem, respectively, from the coherent displacement and from the squeezing of the vacuum state. Specifically, $\mathbf{P}_\text{cl}(\rr) = \lim_{\rho\to0}[\mathbf{P}(\rr)-\mathbf{P}_\text{vac}(\rr)]$ is the Poynting vector associated to a coherent state and $\mathbf{P}_\text{sq}(\rr) = \lim_{\alpha\to0}[\mathbf{P}(\rr)-\mathbf{P}_\text{vac}(\rr)]$ is the Poynting vector associated to a squeezed vacuum state. They are given by
\begin{equation} \label{eq:Scl}
\frac{\mathbf{P}_\text{cl}(\rr)+\mathbf{P}_\text{sq}(\rr)}{\hbar c^2} =
-(|\alpha|^2+\rho^2)\text{Im}[\mathbf{F}_{\kappa_0}(\rr)\times \nabla\times \mathbf{F}_{\kappa_0}^*(\rr)].
\end{equation}
Note that the Poynting vector has the same spatial distribution for both a coherent state and a squeezed vacuum state.

Let us now focus on the contribution stemming from the classical part of the state, $\mathbf{P}_\text{cl}$, and show how to recover the classical scattering cross section obtained within Lorenz-Mie theory~\cite{Lorenz1890,Mie1908}.  First, note that in the absence of the sphere \eqnref{eq:Scl} leads to
\begin{align}
\lim_{\epsilon\rightarrow 1}\mathbf{P}_\text{cl}(\rr) &=  \hbar c^2|\alpha|^2\text{Im}[ \mathbf{G}_{\kappa_0}(\rr)\times \text{i}\kk\times \mathbf{G}_{\kappa_0}^{*}(\rr)]\label{eq:I0defline1} \\
&= \frac{c  |\alpha|^2 \hbar\omega_0}{(2\pi)^3}\uv_k\equiv I_0\uv_k,
\end{align}
where $\omega_0= \omega_{\kappa_0}$ and $\uv_k = \kk / |\kk|$. This expression allows us to define the incoming intensity $I_0$ of the electromagnetic field. On the other hand, the classical scattering cross section, which quantifies the scattered power in the far-field in terms of the incoming intensity, is given by
\begin{align}
\sigma_\text{cl}&\equiv \lim_{|\rr|\rightarrow \infty} \frac{|\rr|^2}{I_0} \int_{\mathbb{S}^2}\text{d}\Omega\,\mathbf{P}_\text{sc}(\rr)\cdot \uv_r,
\end{align}
where $\mathbf{P}_\text{sc}(\rr)$ denotes the Poynting vector for the scattered part of the outgoing normalized scattering modes in Mie form (\eqnref{eq:mie_form})~\cite{Bohren2004}, and reads
\begin{equation}\notag
\mathbf{P}_\text{sc}(\rr) = -\hbar c^2|\alpha|^2 \text{Im}\sum_{\alpha\alpha'}d_{\alpha \kappa_0}d^*_{\alpha'\kappa_0}\mathbf{S}^{\text{sc}}_\alpha(\rr)\times[\nabla\times\mathbf{S}^{\text{sc}}_{\alpha'}(\rr)]^*.
\end{equation}
Using the far-field expression for the scattered part of the eigenmodes (see \secref{sec:sphericalmodes}) and the orthonormality relations of the vector spherical harmonics, we obtain the following expression for the classical scattering cross section,
\begin{equation}\label{eq:squeezingscatteringcrosssection}
\sigma_\text{cl}=\frac{16 \pi^2}{|\kk|^2}\sum_{plm}|c_{lmg}^p\sin\varphi_l^p|^2=\sigma.
\end{equation}
This expression is equal to the scattering cross section obtained within Lorenz-Mie theory~\cite{Bohren2004} and the single-photon scattering cross section \eqnref{eq:single_photon_scattering_cross_section} in \secref{sec:onephoton}.

Let us briefly comment on the contribution of the squeezing to the scattering cross section. Using the definition of the vectors $\mathbf{P}_\text{cl}(\rr)$ and $\mathbf{P}_\text{sq}(\rr)$ and \eqnref{eq:Scl}, it is straightforward to show that the derivation in Eqs.~\eqref{eq:I0defline1} to \eqref{eq:squeezingscatteringcrosssection} also holds for a coherent and squeezed state under the substitution $\vert\alpha\vert^2 \rightarrow \vert\alpha\vert^2 + \rho^2$. In particular, the scattering cross section \eqnref{eq:squeezingscatteringcrosssection} remains unchanged and is thus independent on the degree of squeezing of the incoming beam or on the nature of such squeezing (amplitude, phase, or both).
Note that, although squeezing does not affect the angular distribution of the intensity scattered by a sphere, it might modify other quantities such as electromagnetic forces (e.g. dipole or scattering force) exerted on a sphere that is allowed to move. The study of these more complex processes is left for future work. 
In general, a rigorous characterization of the scattering of squeezed light is only possible through the quantum theoretical formulation derived in this work.

\subsection{Two-photon state: Hong-Ou-Mandel effect}
\label{sec:twophoton}

Let us now consider a situation where observable quantum effects arise, namely the scattering of a two-photon state off a dielectric sphere. By using the sphere as a beam splitter, we aim at finding situations where the simultaneous joint detection probability of the two scattered photons vanishes due to quantum interference, a phenomenon known as Hong-Ou-Mandel interference~\cite{Hong1987}.
To this aim, the quantity of interest is the joint probability to detect, at a given time $t$, a $i$-polarized photon at a position $\rr_1$ and a $j$-polarized photon at a position $\rr_2$. This probability is proportional to the second-order correlation function \cite{Glauber1963}
\begin{equation}\label{eq:G2}
G^2_{ij}(\rr_1,\rr_2,t)=\langle \hat{E}_i^-(\rr_1,t)\hat{E}_j^-(\rr_2,t)\hat{E}_j^+(\rr_2,t)\hat{E}_i^+(\rr_1,t)\rangle,
\end{equation}
where $\hat{E}^+_i(\rr,t)=[\hat{\mathbf{E}}^+(\rr,t)]\cdot\uv_i=[\hat{E}^-_i(\rr,t)]^\dagger$ is the projection of the positive frequency part (terms with annihilation operators) of the electric field operator in the Heisenberg picture on the polarization vector. It is useful to also define the normalized second-order correlation function
\begin{equation}
\label{eq:g2_not_evaluated}
g^2_{ij}(\rr_1,\rr_2,t)=\frac{G^2_{ij}(\rr_1,\rr_2,t)}{G^1_i(\rr_1,t)G^1_j(\rr_2,t)},
\end{equation}
defined in terms of the second-order correlation function  and the first-order correlation function $G_i^1(\rr,t)=\langle \hat{E}_i^-(\rr,t) \hat{E}_i^+(\rr,t)\rangle $. 

Let us now evaluate $g^2_{ij}(\rr_1,\rr_2,t)$ for a two-photon state. Each photon is assumed to populate an outgoing normalized scattering eigenmode, with respective mode indices $\kappa_1$ and $\kappa_2$. 
To avoid a non-normalizable (Dirac delta) representation of the total state, we can write it as
\begin{equation}\label{eq:initial2photonstate}
\ket{\psi}= \sum_{\kappa \kappa'}\phi_{\kappa\kappa_1}\phi_{\kappa'\kappa_2} (\hat{a}_\kappa^\text{out})^\dagger(\hat{a}_{\kappa'}^\text{out})^\dagger \ket{0},
\end{equation}
where the vacuum state is defined with respect to the same eigenmodes $\hat{a}_{\kappa}^\text{out}\ket{0}=0$. The weight functions $\phi_{\kappa\kappa_n}$ associated with mode $\kappa_n$ are not explicitly specified but they are assumed to fulfill three properties: (i) they are normalized, i.e  $\sum_{\kappa}|\phi_{\kappa \kappa_n}|^2=1$, (ii) the modes are assumed to be sufficiently distinguishable, such that $\sum_{\kappa}\phi^*_{\kappa \kappa_1}\phi_{\kappa \kappa_2}\simeq 0$, and (iii) the weight functions are sufficiently peaked around $\kappa_n$ such that  $\sum_{\kappa}\phi_{\kappa \kappa_n}f_{\kappa}\propto f_{\kappa_n}$ for any sufficiently well-behaved function $f_{\kappa_n}$. As a remark, note that the correlation functions \eqnref{eq:G2} and \eqnref{eq:g2_not_evaluated} are defined in the Heisenberg picture, where quantum states do not evolve in time. Hence, by virtue of \eqnref{eq:id1} and \eqnref{eq:id2}, the correlation functions obtained for the state \eqnref{eq:initial2photonstate} and for the two-photon \textit{plane wave} state labelled by the indices $\kappa_1$ and $\kappa_2$ would be identical.
For the state \eqnref{eq:initial2photonstate}, the normalized second-order correlation function can be readily evaluated using properties (i)-(iii) and reads
\begin{equation}
g_{ij}^{2}(\rr_1,\rr_2)=
\frac{\omega_{\kappa_1}\omega_{\kappa_2}|F_{i\kappa_1}(\rr_1)F_{j\kappa_2}(\rr_2)+F_{i\kappa_2}(\rr_1)F_{j\kappa_1}(\rr_2)|^2}{\sum_{n=1,2}\omega_{\kappa_n}|F_{i\kappa_n}(\rr_1)|^2\sum_{n=1,2}\omega_{\kappa_n}|F_{j\kappa_n}(\rr_2)|^2},
\end{equation}
where $F_{i\kappa}(\rr)= \mathbf{ F}^\text{out}_{\kappa}(\rr)\cdot \uv_i$ is the projection of the outgoing normalized scattering eigenmode on the polarization vector. Note that the numerator accounts for the superposition -- interference -- of two processes: either the photon in mode $\kappa_1$ reaches the detector at $\rr_1$ while the photon in mode $\kappa_2$ reaches the detector at $\rr_2$ or vice versa.

\begin{figure}
	\centering
	\includegraphics[width=0.5\textwidth]{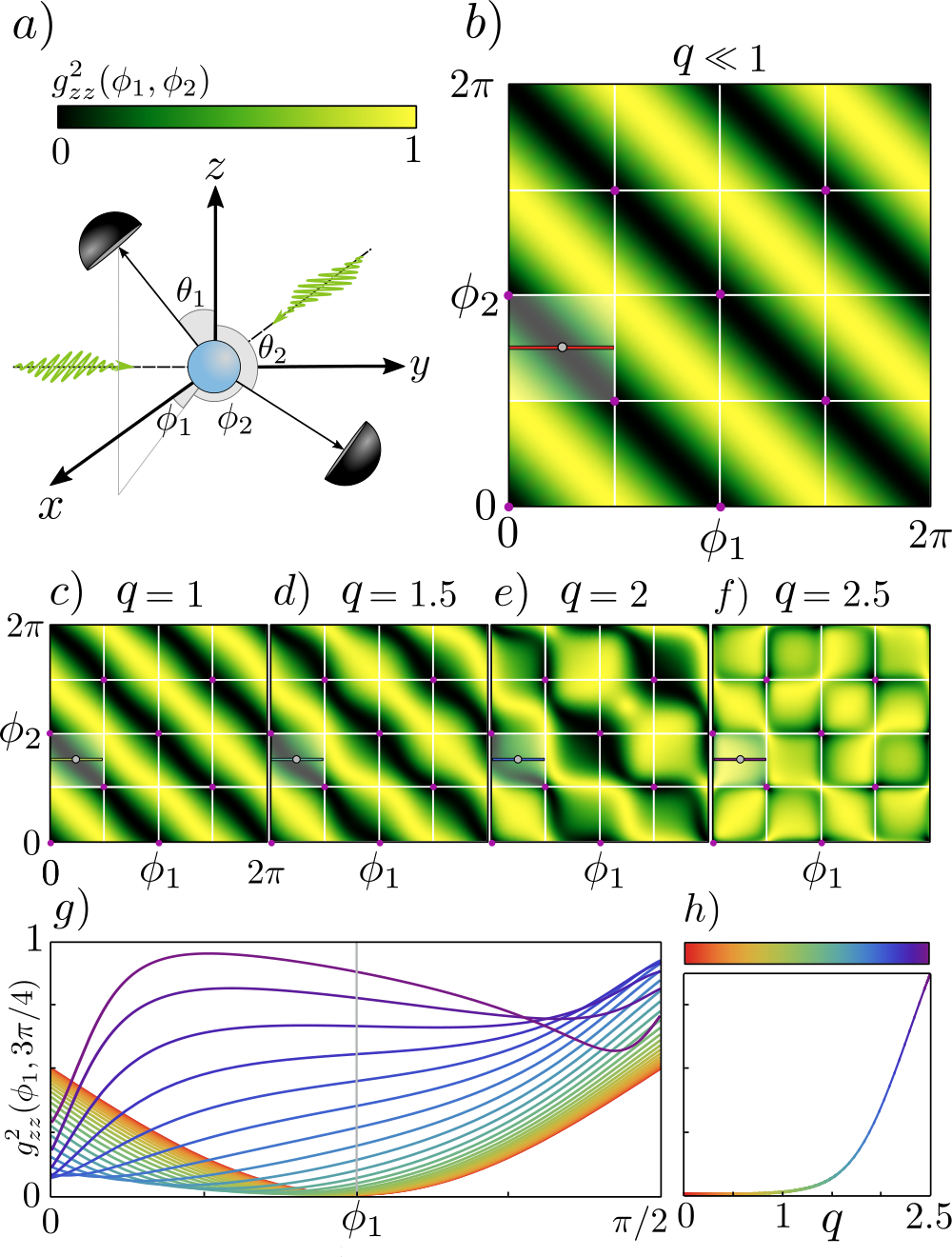}
	\caption{(a) Illustration of the two-photon state and the detector's positions in relation to the dielectric sphere (not to scale). The two photons occupy the modes $\kappa_1=(1,\omega_{\kappa}\uv_x/c)$ and $\kappa_2=(1,\omega_{\kappa}\uv_y/c)$ which corresponds to a normalized scattering eigenmode with a $y(x)$-polarized plane-wave part travelling along the positive $x(y)$-axis. (b)-(f) Normalized second-order correlation function of the $z-$component of the electric field, $g^2_{zz}(\phi_1,\phi_2)$, for $\theta_1 =\theta_2-\pi/2= \pi/4$ as a function of the azimuthal positions of the two detectors and for different values for $q=kR$. The purple dots show the points for which $g^2_{zz}(\phi_1,\phi_2)=0$ because of the fact that one of the two photons does not induce a signal in neither of the two detectors. (g) Normalized second-order correlation function $g^2_{zz}(\phi_1,\phi_2=3\pi/4)$ for different values of $q$. (h) Normalized second-order correlation function $g^2_{zz}(\phi_1,\phi_2)$ for $\phi_1+\pi=\phi_2=3\pi/4$  and different values of $q$. The color scales of panels (g) and (h) are the same.}
	\label{fig:1}
\end{figure}

Let us characterize the second-order correlation function for the particular example depicted in \figref{fig:1}(a). We consider two modes of equal frequency $\omega_{\kappa_n}=ck$, the first with an $y-$polarized plane wave part and travelling along the positive $x-$axis, and the second with an $x-$polarized plane wave part and travelling along the positive $y-$axis. These modes correspond to $\kappa_1=(1,k\uv_x)$ and $\kappa_2=(1,k\uv_y)$. Moreover, we assume the detectors to be placed at a fixed distance $|\rr_1|=|\rr_2|\gg R$ in the far-field of the sphere. The position of each detector is therefore fully determined by their polar and azimuthal angles $(\theta_i,\phi_i)$, with $i=1,2$. Subsequently, we fix their polar position $\theta_1 =\theta_2-\pi/2= \pi/4$. For small particles the far-field correlation function is dominated by the plane-wave part rather than the scattered part unless the detector is not susceptible to the polarization of the plane-wave part. Thus, in order to study the correlation function of the scattered part we assume the detectors to be solely susceptible to $z$-polarized photons ($i=j=z$).

The function $g^2_{zz}(\phi_1,\phi_2)$ is shown in \figref{fig:1}(b-f) as a function of the detector's azimuthal positions $\phi_1$ and $\phi_2$, for different values of $q$, and for a sphere with relative permittivity  $\epsilon = 2.1$. Convergence of the plotted expressions has been checked numerically. \figref{fig:1}(b) corresponds to the normalized second-order correlation function in the small-particle limit. The purple dots show the points for which $g^2_{zz}(\phi_1,\phi_2)=0$ due to one of the two photons inducing a signal never arriving to one of the two detectors. This case is discussed in more detail later. All other pairs $(\phi_1,\phi_2)$ resulting in a vanishing correlation function (black diagonal stripes in \figref{fig:1}(b)) correspond to Hong-Ou-Mandel destructive interference. As can be seen in \figref{fig:1}(c)-(f) the correlation functions change appreciably as $q$ increases while the regions of interference, $g^2_{zz}(\phi_1,\phi_2)\simeq 0$, shrink. \figref{fig:1}(g)-(h) show a section of the normalized second order correlation function (marked by thick horizontal segments in \figref{fig:1}(b-f)) for different values of $q$.  For a fixed $\phi_1$ the correlation function assumes a broad range of different values throughout $0<q\leq 2.5$. Specifically, for small particles this function vanishes, certifying, as we discuss below, the quantum nature of the interference.

To better understand the different implications of a vanishing normalized second-order correlation function, let us focus on the small-particle limit, for which an analytical expression can be derived. Specifically, the scattered part of the scattering eigenmodes $\kappa_1$ and $\kappa_2$ in Mie form (\eqnref{eq:mie_form}) can be expanded in orders of $q$. In the small-particle regime $q\ll 1$ the dominant contribution in the far-field reads
\begin{align}\label{eq:ScModessmallparticlelimit}
\lim_{|\rr|\rightarrow \infty} F_{z\kappa_1}(\rr)& =F_0\frac{q^3\exp(\text{i} k r)}{2k r}\sin(2\theta)\sin\phi,\\
\lim_{|\rr|\rightarrow \infty} F_{z\kappa_2}(\rr)& =F_0^*\frac{q^3\exp(\text{i} k r)}{2k r}\sin(2\theta)\cos\phi,
\end{align}
with
\begin{equation}
    F_0= \frac{-\text{i} }{(2\pi)^{3/2}}\frac{\epsilon-1}{\epsilon+2}.
\end{equation}
Inserting these expressions in \eqnref{eq:g2_not_evaluated}  we obtain
\begin{equation}
\label{eq:g2_small_particle}
g_{zz}^{2}(\rr_1,\rr_2)\simeq \sin^2(\phi_1+\phi_2).
\end{equation}
Note that, according to Eqs.~\eqref{eq:ScModessmallparticlelimit}, no $z-$polarized photons are detected for $\theta_1,\theta_2\in \lbrace 0,\pi/2,\pi\rbrace$, as the $z$ component of the mode function vanishes for both photons. 
While the second-order correlation function $G_{ij}^{2}(\rr_1,\rr_2)$ vanishes for these values, this is not the case for $g_{ij}^{2}(\rr_1,\rr_2)$ due to the normalization. All remaining pairs of coordinates $(\phi_1,\phi_2)$ for which a vanishing normalized second-order correlation function equals a vanishing joint detection probability can be subdivided into two categories. The pairs for which either $\sin\phi_1=\sin\phi_2=0$ or $\cos\phi_1=\cos\phi_2=0$ lead to a vanishing correlation function because one of the two photons has a zero detection probability at both detectors (see again Eqs.~\eqref{eq:ScModessmallparticlelimit}). All other pairs, namely the pairs fulfilling $\phi_1+\phi_2= n\pi$ with $n\in\mathbb Z$, lead to a vanishing correlation function -- and thus vanishing joint detection probability -- due to destructive interference, namely $F_{z\kappa_1}(\rr_1)F_{z\kappa_2}(\rr_2)=-F_{z\kappa_2}(\rr_1)F_{z\kappa_1}(\rr_2)$ while $F_{z\kappa_1}(\rr_1)F_{z\kappa_2}(\rr_2)\neq 0$. We remark that for two classical (coherent) states in modes $\kappa_1$ and $\kappa_2$ the joint detection probability $g_{ij}^{2}(\rr_1,\rr_2)$ would also display minima corresponding to destructive interference, but it would not vanish at these minima. A vanishing correlation function $g_{ij}^{2}(\rr_1,\rr_2)=0$ at these points is a signature of the quantum nature of the two incoming single-photon states, and is known as Hong-Ou-Mandel interference. Finally, we would like to emphasize that the features of $g^2_{zz}(\rr_1,\rr_2)$ strongly depend on the refractive index, size and the shape of the sphere. This could render the normalized second-order correlation function a sensitive tool to perform tomography of a polarizable object.

\subsection{Two-photon state: Scattering Amplitude and Cross Section} 
\label{sec:2p_transition_amplitude}
Let us now study the scattering of a two-photon plane wave state by deriving the two-photon scattering amplitude and cross section. In analogy to the single-photon case described in \secref{sec:onephoton}, the asymptotic two-photon transition amplitude from the initial state $\hat{b}_{\kappa'_1}^\dagger\hat{b}_{\kappa'_2}^\dagger \ket{0}$ to the final state $\hat{b}_{\kappa_1}^\dagger\hat{b}_{\kappa_2}^\dagger \ket{0}$ reads
\begin{align}
\mathcal{T}_{\kappa_1'\kappa_2'}^{\kappa_1\kappa_2} &=\bra{0} \hat{b}_{\kappa_1}\hat{b}_{\kappa_2} \hat{U}(t_+,t_-) \hat{b}^\dagger_{\kappa_1'}\hat{b}^\dagger_{\kappa_2'}  \ket{0}=S_{\kappa_1'\kappa_2'}^{\kappa_1\kappa_2}\exp[-\text{i} (\omega_{\kappa_1}+\omega_{\kappa_2})(t_+-t_-)].
\end{align}
Using \eqnref{eq:asymptotic_transformation} the resulting two-photon scattering matrix can be shown to read
\begin{equation}
\label{eq:two_photon_scattering_matrix}
S_{\kappa_1'\kappa_2'}^{\kappa_1\kappa_2}=S_{\kappa_1\kappa_1'}S_{\kappa_2\kappa_2'}+S_{\kappa_1\kappa_2'}S_{\kappa_2\kappa_1'},
\end{equation}
where $S_{\kappa\kappa'}\propto \delta(|\kk|-|\kk'|)$ denotes the single-photon scattering matrix, for which the explicit expression is given in \eqnref{eq:scatt_mat}. The fact that \eqnref{eq:two_photon_scattering_matrix} is written as a sum of terms preserving the individual energy of each photon, implies that the same is true for the two-photon scattering process. This is in contrast to nonlinear scattering processes, where the individual energy of each photon is not conserved, but the total energy is. 

We further evaluate \eqnref{eq:two_photon_scattering_matrix} by assuming that the initial states are monochromatic with wave number $|\mathbf{k}|$. Inserting \eqnref{eq:scatt_mat} we arrive at the following expression
\begin{align}\notag
S_{\kappa_1'\kappa_2'}^{\kappa_1\kappa_2} &= \delta_{\kappa_1\kappa_1'}\delta_{\kappa_2\kappa_2'} + \delta_{\kappa_1\kappa_2'}\delta_{\kappa_2\kappa_1'}\\\notag
&+\frac{\text{i}}{2\pi |\kk|}\left(\delta_{\kappa_1\kappa_1'} f_{\kappa_2\kappa_2'}+\delta_{\kappa_1\kappa_2'}f_{\kappa_2\kappa_1'} \right)\delta(|\kk_2|-|\kk|)\\\notag
&+\frac{\text{i}}{2\pi |\kk|}\left(\delta_{\kappa_2\kappa_2'} f_{\kappa_1\kappa_1'}+\delta_{\kappa_2\kappa_1'}f_{\kappa_1\kappa_2'} \right)\delta(|\kk_1|-|\kk|)\\
&-\frac{1}{4\pi^2 |\mathbf{k}|^2}f_{\kappa_1'\kappa_2'}^{\kappa_1\kappa_2}\delta(|\kk_1|-|\mathbf{k}|)\delta(|\kk_2|-|\mathbf{k}|),
\end{align}
where the first line accounts for all processes in which both photons are unaffected by the presence of the sphere, the second and third line account for all single-photon scattering processes, and the resulting two-photon scattering amplitude reads $f_{\kappa_1'\kappa_2'}^{\kappa_1\kappa_2}\equiv f_{\kappa_1\kappa_1'}f_{\kappa_2\kappa_2'}+f_{\kappa_1\kappa_2'}f_{\kappa_2\kappa_1'}$, where $f_{\kappa\kappa'}$ denotes the single-photon scattering amplitude. This allows us to define the two-photon scattering cross section~\cite{Cohen2004/2}
\begin{equation}
    \sigma^{(2)}_{\kappa_1'\kappa_2'}= \frac{1}{2} \sum_{g_1 g_2} \int \text{d}\Omega_{k_1}\text{d}\Omega_{k_2}|f_{\kappa_1\kappa_1'}f_{\kappa_2\kappa_2'}+f_{\kappa_1\kappa_2'}f_{\kappa_2\kappa_1'}|^2.
\end{equation}
Inserting \eqnref{eq:scattering_amplitude_elastic} for the single-photon scattering amplitude and using the orthonormality relations of the vector spherical harmonics we can further simplify the above expression and arrive at
\begin{align}
    \sigma^{(2)}_{\kappa_1'\kappa_2'}= \sigma^2+  \left| \frac{16\pi^2}{|\mathbf{k}|^2}\sum_{plm} c_{lmg_1'}^{p*} c_{lmg_2'}^{p} \sin ^2\varphi_l^p\right|^2.
\end{align}
Here, $\sigma$ denotes the single-photon scattering cross section as defined in \eqnref{eq:definition_differential_scattering_cs}. This shows that the two-photon scattering cross section is lower bounded by $\sigma^2$. As an example, let us particularize to the small-particle limit with $g'_{1}=g'_{2} = 1$. Up to lowest order in $q=|\kk|R$ we arrive at
\begin{align}
    \sigma^{(2)}_{\kappa_1'\kappa_2'}=\sigma^2[1 + \cos^2(\phi_{\kappa_1'}-\phi_{\kappa_2'
    })],
\end{align}
where the explicit expression for the single-photon scattering cross section in the small-particle limit is given in \eqnref{eq:scattering_cross_section_elastic_small}. We see that in the small particle-limit the two-photon scattering cross section is proportional to the square of the single-photon scattering cross section and reaches its maximum value for $\phi_{\kappa_1'}=\phi_{\kappa_2'}$.

\section{Conclusions}
\label{sec:conclusions}

In this work we have taken the first step toward understanding the interaction between light and a dielectric sphere by developing a quantum theory of light scattering off a nonmoving and rigid dielectric sphere, i.e., a sphere without dynamical degrees of freedom. Following the seminal derivation of Glauber and Lewenstein~\cite{Glauber1991}, we have quantized the electromagnetic field in the presence of the sphere in terms of normalized eigenmodes and derived the canonical transformation between such eigenmodes and plane waves. We have extended their work and applied it to a spherical geometry, whose symmetry has allowed us to analytically solve the scattering of a single-photon state, a coherent and squeezed state, and a two-photon state for spheres of arbitrary size. These three relevant examples illustrate the potential of our quantum framework to rigorously describe the scattering of non-classical states of light.

More importantly for levitated optomechanics~\cite{MillenRepProgPhys2020,GonzalezBallestero2021}, this article sets the basis to develop an extended theoretical formalism including relevant dynamical degrees of freedom of the dielectric sphere, such as motion as done by us in~\cite{Maurer2023}, rotation, or acoustic vibrations. As discussed in~\cite{Maurer2023}, this must involve, first, a derivation of the  electromagnetic field Hamiltonian in the presence of a dynamical dielectric medium (allowed to e.g. move or vibrate), and second,  the consideration of new, \textit{inelastic} scattering processes that typically arise when scatterers have external (e.g. center-of-mass motion) and internal degrees of freedom (e.g. Brillouin scattering with acoustic phonons).
Our work provides the tools to rigorously address both aspects. Extending our formalism to include the dynamical degrees of freedom of a sphere is crucial to precisely derive quantities as relevant as light-matter interaction coupling rates or heating rates due to photon recoil~\cite{Maurer2023}, as well as to open new directions for controlling levitated dielectric spheres~\cite{LepeshovPRL2023}.

\begin{backmatter}
\bmsection{Funding}
This research was supported by the European Union’s Horizon 2020 research and innovation programme under grant agreement No. [863132] (IQLev) and from the European Research Council (ERC) under the grant Agreement No. [951234] (Q-Xtreme ERC-2020-SyG).

\bmsection{Acknowledgments}
We acknowledge valuable discussions with Daniel H\"ummer.

\bmsection{Disclosures}
The authors declare no conflicts of interest.

\bmsection{Data availability} Data underlying the results presented in this paper are not publicly available at this time but may be obtained from the authors upon reasonable request.

\end{backmatter}

\bibliography{references}

\end{document}